\documentclass[12pt,letterpaper,a4paper]{article}
\usepackage[includeheadfoot,
            marginratio={1:1,2:3},
            width=412pt,
            height=688pt,]{geometry}
\usepackage{amsmath}
\usepackage{amsfonts}
\usepackage{amssymb}
\usepackage{graphicx}
\usepackage{empheq}
\usepackage{subfigure}
\usepackage{color}
\usepackage{hyperref}
\usepackage{enumerate}

\newcommand{\nc}{\newcommand}
\nc{\beq}{\begin{equation}}
\nc{\eeq}{\end{equation}}
\nc{\bea}{\begin{eqnarray}}
\nc{\eea}{\end{eqnarray}}
\def\ov{\overline}
\def\cO{{\cal O}}
\def\IP{\mathbb{P}}
\def\f{\frac}

\newdimen\csize\csize=1.5ex
\def\young#1{\tiny\vcenter{\hbox{\vrule\vtop{\hrule
  \offinterlineskip\halign{&\vbox
  {\hbox to\csize {\strut\hss##\hss\vrule}\hrule}\cr#1 \crcr}}}}}

\newcommand{\eq}[1]{\begin{equation}
                     \begin{split} #1 \end{split}
                     \end{equation}}

\begin{document}

\vspace*{-1.5cm}
\begin{flushright}
  {\small
  MPP-2013-15\\
  }
\end{flushright}

\vspace{1.5cm}
\begin{center}
  {\Large
On Non-Gaussianities in Two-Field \\ 
Poly-Instanton Inflation}

\end{center}

\vspace{0.75cm}
\begin{center}
Xin Gao$^{\dagger, \,  \ddagger}$\footnote{Email: gaoxin@mppmu.mpg.de} and Pramod Shukla$^\dagger$\footnote{Email: shukla@mppmu.mpg.de}
\end{center}

\vspace{0.1cm}
\begin{center}
\emph{
$^{\dagger}$ Max-Planck-Institut f\"ur Physik (Werner-Heisenberg-Institut), \\
   F\"ohringer Ring 6,  80805 M\"unchen, Germany\\
\vskip 1cm
$^{\ddagger}$ State Key Laboratory of Theoretical Physics, \\
Institute of Theoretical Physics,\\ Chinese Academy of Sciences, P.O.Box 2735, Beijing 100190, China } \\[0.1cm]
\vspace{0.2cm}

 \vspace{0.5cm}
\end{center}

\vspace{1cm}


\begin{abstract}
In the context of Type IIB LARGE volume orientifold setup equipped with poly-instanton corrections, the standard single-field poly-instanton inflation driven by a `Wilson' divisor volume modulus is generalized by the inclusion of respective axion modulus. This two-field dynamics results in a ``Roulette" type inflation with the presence of several inflationary trajectories which could produce 50 (or more) e-foldings. The  evolution of various trajectories along with physical observables are studied. The possibility of generating primordial non-Gaussianities in the slow-roll as well as in the beyond slow-roll region is investigated. We find that although the non-linearity parameters are quite small during the slow-roll regime, the same are significantly enhanced in the beyond slow-roll regime investigated up to the end of inflation.
\end{abstract}

\clearpage



\section{Introduction}
On the way of understanding the early universe cosmology, the inflationary mechanism has been proven to be quite fascinating as it successfully addresses several outstanding issues of the standard big bang scenario, e.g. the horizon problem, the flatness problem and the monopole problem. Initially, the idea of inflation was introduced to explain the homogeneous and isotropic nature of the universe at large scale structure \cite{Guth:1980zm,Linde:1981mu}, however its best advantage is being utilized in studying the inhomogeneities and anisotropies of the universe, which is a consequence of the vacuum fluctuations of the inflaton (as well as the metric fluctuations). In this regard, the idea of inflation also provides a way to understand the physics which could be responsible for generating the correct amount of primordial density perturbations initiating the structure formation of the universe and the cosmic microwave background (CMB) anisotropies. The simplest (single-field) inflationary process can be 
understood via a (single) scalar field slowly rolling towards its in a nearly flat potential. The vacuum fluctuations of the inflaton result in an almost scale invariant spectrum with a small tilt reflecting unique predictions via the CMB radiation. Although the single-field inflationary models fit well with the current observation constraints \cite{Larson:2010gs,Komatsu:2010fb},
the present observational data is not sufficient to discriminate among the various models. In this regard, the detection of non-Gaussianity can be a crucial data to distinguish the various models in the ongoing/future experiments such as PLANCK \cite{Bartolo:2004if,Planck:2006aa,Yadav:2007yy,Efstathiou:2009xv}. 

In the context of string cosmology, inflationary model building have been started quite early in \cite{Dvali:1998pa}. 
Since string framework can provide several flat-directions (moduli), it is promising for the embedding of inflationary scenarios in string theory. 
The main requirement for this purpose is to identify a scalar which could
play the role of the inflaton field, i.e.\ its effective scalar potential has
to admit a slow-roll region. In this respect, those ``moduli'' that have a
flat potential at leading order and only by a sub-leading effect receive
their dominant contribution are of interest.
The perturbative effects \cite{Becker:2002nn,Gukov:1999ya} as well as the instanton effects \cite{Witten:1996bn,Blumenhagen:2009qh} are proven to be extremely
crucial,F especially for moduli stabilization purpose.
As far as cosmological model building is concerned, the string inspired models could be given the hallmark of being `(semi)realistic' only after all the moduli could be stabilized and de-Sitter solutions be realized in \cite{Kachru:2003aw}. Since then several dS realizing mechanisms have been developed in type IIB framework \cite{Burgess:2003ic,Saltman:2004sn,Westphal:2006tn,Misra:2007yu,Cicoli:2012fh,Louis:2012nb}.
With the present understanding, it is fair to say that the moduli stabilization is quite well (and relatively much better) 
understood in type IIB orientifold models with the mechanisms like KKLT \cite{Kachru:2003aw}, Racetrack \cite{BlancoPillado:2004ns,Abe:2005pi} and the LARGE volume scenarios \
\cite{Balasubramanian:2005zx}. Significant amount of progress has been made in building up inflationary models in type IIB orientifold setups with the inflaton field identified as an open string modulus \cite{Kachru:2003sx,Dasgupta:2004dw,Avgoustidis:2006zp,Baumann:2009qx} or a closed string modulus \cite{Conlon:2005jm,Conlon:2008cj,Cicoli:2008gp,Cicoli:2011zz,Blumenhagen:2012ue}. 
Along the lines of moduli getting lifted by  sub-dominant contributions,
recently so-called poly-instanton corrections became  of interest.
These are  sub-leading non-perturbative contributions which can be briefly
described as instanton corrections to instanton actions. 
In the recent work \cite{Blumenhagen:2012kz}, we have clarified
the zero mode conditions for an Euclidean D3-brane instanton, wrapping a
divisor of the threefold, to generate such a poly-instanton effect.
Utilizing the  poly-instanton effects, moduli stabilization and inflation
have been studied in a series of papers
\cite{Blumenhagen:2012ue,Blumenhagen:2008kq,Cicoli:2011ct,Cicoli:2012cy}.
Meanwhile, the studies made in the context of axionic inflationary models in the type IIB orientifold framework have been quite promising too \cite{BlancoPillado:2004ns,Dimopoulos:2005ac,BlancoPillado:2006he,Kallosh:2007cc,Cicoli:2012tz,Grimm:2007hs,Misra:2007cq,McAllister:2008hb}. In \cite{Burgess:2008ir}, inflation has been realized by a combination of the brane-motion and an axion while in \cite{Kallosh:2004rs,Bond:2006nc,BlancoPillado:2009nw},, it has been driven by a combination of a (divisor) volume mode and an axion.

The signatures of non-Gaussianities are encoded in non-linearity parameters $f_{NL}$ (and $\tau_{NL}$) which parametrize the bispectrum (and trispectrum) and could be the direct evidence for detecting the  non-Gaussianities. In fact, an observation of order one values for $f_{NL}$ would rules out (most of) the single field inflationary models \footnote{There has been some proposals of single-field models with large $f_{NL}$ values \cite{Qiu:2010dk, Namjoo:2012aa, Noller:2011hd, Engel:2008fu}.}. 
and thus can be a potential discriminator towards picking up a more promising one. 
This boosts up the motivation for the study of non-Gaussianties in recent years, although it has been initiated very early \cite{Maldacena:2002vr}. A review on initial attempts regarding $f_{NL}$ computations along with observational constraints can be found in \cite{Bartolo:2004if}. In the standard single-field models, the $f_{NL}$ parameter is usually suppressed by slow-roll parameters ($\epsilon, \eta$), e.g. $f_{NL} = n_s -1$ has been reported in \cite{Maldacena:2002vr}. Subsequently, the multi-field inflationary models started getting significant attention but large $f_{NL}$ 
could not be realize in the initial attempts \cite{Vernizzi:2006ve,Battefeld:2006sz,Choi:2007su,Yokoyama:2007uu}. For generating observable non-Gaussianities, enormous amount of work has been done in recent couple of years and several explicit multi-field models with large $f_{NL}$ values have been found. These multi-field models  can be categorized to be of separable and non-separable types.

For the separable type models, either the inflationary potential \cite{Vernizzi:2006ve,Battefeld:2006sz,Choi:2007su,Rigopoulos:2005us,Seery:2006js} or the Hubble rate \cite{Byrnes:2009qy,Battefeld:2009ym} is of sum or product separable form, and several models are available now which can produce large non-Gaussianities \footnote{A nice recent review about generating large non-Gaussianities for separable type setups can be found in \cite{Byrnes:2010em,Suyama:2010uj}.}. The possibilities of generating detectable large values of $f_{NL}$ during the evolution of two-fields within the slow-roll region has been proposed in \cite{Byrnes:2008wi,Byrnes:2008zz,Byrnes:2008zy}. In these models, non-Gaussianities are generated after horizon exit and consist of non-adiabatic perturbation modes. 
It implies such non-Gaussianities to be of local-type and hence distinguishable from the other shapes of non-Gaussianity which are realized during horizon exit \cite{Komatsu:2009kd,Fergusson:2008ra}.

Unlike the separable type models, the non-Gaussianities issues in the non-separable type of potential are relatively less studied.
In \cite{Yokoyama:2007uu,Yokoyama:2007dw,Yokoyama:2008by}, a concise analytic formula for computing the non-linear parameter for a given {\it generic } multi-field potential has been developed for both the slow-roll region and beyond slow-roll region. The possibility of generating large non-Gaussianity parameter has been shown by `momentarily' violating the slow-roll conditions.
Following the ideas developed in \cite{Yokoyama:2007dw}, order one values of the $f_{NL}$ parameter has been reported in the beyond slow-roll region for a two-field large volume axionic setup \cite{Misra:2008tx}. Recently, some examples with (non-)separable multifield potentials have been studied in \cite{Mazumdar:2012jj} which can produce large detectable values for the non-linear parameter $f_{NL}$ and $\tau_{NL}$. Higher order correction to the non-linearity parameters have been studied in \cite{Lyth:2005fi,Zaballa:2006pv,Cogollo:2008bi,Rodriguez:2008hy}. 

In the framework of type IIB orientifolds, several single/multi-field models have been studied for aspects of non-Gaussianities. The curvaton scenarios have been proposed in \cite{Burgess:2010bz} for the K\"ahler moduli inflation setup in the context of LARGE volume scenarios. A case study have been performed in \cite{Berglund:2010xr} for a blow-up multi-field inflationary setup (resulting in $f_{NL} \sim 0.01$) and a curvaton scenario (resulting in large $f_{NL}$ values). Later on, detectable large values of $f_{NL}$ have also been proposed via the modulated reheating mechanism in \cite{Cicoli:2012cy}. The computation of non-Gaussianties in racetrack models has been made in \cite{Sun:2006xv} and within the framework of KKLT-like setup, a two-field inflationary model has been proposed with inflaton dynamics governed by the Calabi Yau volume mode and the respective $C_4$ axion which complexifies the divisor volume mode \cite{Kallosh:2004rs}. This idea has been extended in the context of LARGE volume scenarios 
by generalizing 
the  K\"ahler moduli (blow-up) inflation with inclusion of the corresponding $C_4$ axion in \cite{Bond:2006nc,BlancoPillado:2009nw}. The presence of various types of trajectories resulting in 60 (or more) number of e-foldings have been reported in this so-called ``Roulette inflation'' model and a subsequent investigation of non-Gaussianities in such a setup has been made in \cite{Vincent:2008ds} resulting in small values of the $f_{NL}$ parameter.  

In this article, we present a systematic analysis for a two-field inflationary model driven by a combination of a Wilson divisor volume modulus and an axion modulus in the Poly-instanton setup. The so-called Wilson divisor has a single non-trivial Wilson line modulino and is relevant for generating  poly-instanton contributions which will be summarized in Section \ref{sec_Setup}. For analyzing the non-Gaussianities, we proceed with the formalism developed in \cite{Yokoyama:2007dw,Yokoyama:2008by} which is valid for a given generic multi-field potential in beyond slow-roll region also. { We find that although the non-linearity parameters are quite small during the slow-roll regime, the same are significantly enhanced in the beyond slow-roll regime. The two-field dynamics is such that the non-linearity parameters ($f_{NL}, \tau_{NL}$ and $g_{NL}$) do not get frozen at the horizon exit and keep evolving up to the end of inflation where it acquires a large value. Here, we stress that following the motion/decay 
of inflaton fields after the end of inflation and addressing reheating issues etc. can be extremely crucial in such models. However, the same is beyond the scope of the present analysis, and for the time being we assume that the respective values for these non-linearity parameters realized at the end of inflation do not change significantly before and by the reheating process.}

The article is organized as follows. In Section \ref{sec_Setup}, we start with a brief review of the relevant background about the poly-instanton setup \cite{Blumenhagen:2012kz} and the related moduli stabilization part \cite{Blumenhagen:2012ue}. In Section \ref{sec_Inflationary Cosmology}, we generalize the single field poly-instanton inflationary process into a two-field inflationary process with the inclusion of the corresponding $C_4$ axion and discuss the subsequent relevant changes in the slow-roll parameters. In Section \ref{sec_Evolution}, we present a detailed and systematic (numerical) analysis of evolutions of various inflationary trajectories (as well as the other physical observables) in terms of number of e-foldings. This analysis shows that there is a ``Roulette" type inflation. In section \ref{sec_fNL}, we investigate the possibility of generating finite/detectable values for the primordial non-Gaussianity parameters $f_{NL}, \tau_{NL}, g_{NL}$ and observe that although in the slow-roll 
region these are 
negligibly small, beyond it there exists the a possibility of enhancement to large values. Finally, in section \ref{sec_Conclusions and Discussions} we give our conclusions followed by an appendix providing some of the lengthy intermediate expressions.

\section{Poly-Instanton Setup}
\label{sec_Setup}
In this section, we collect the relevant ingredients for generating the poly-instanton corrections in a type IIB orientifold setup developed in \cite{Blumenhagen:2012kz} and briefly summarize the moduli stabilization mechanism discussed in \cite{Blumenhagen:2012ue}. Building on the same, we continue with the study of a generalized two-field inflationary process as well as with the  computation of the non-linearity parameters (e.g. $f_{NL}, \tau_{NL}$ and $g_{NL}$). 

\subsection{Poly-instanton corrections}
\label{sec_Poly-instanton corrections}

Now, we recall some results from \cite{Blumenhagen:2012kz,Blumenhagen:2008ji} (see also \cite{Petersson:2010qu} for related studies)
on the
contribution  of poly-instantons to the superpotential in the
framework of Type IIB  orientifold
compactifications on Calabi-Yau threefolds with $O7$- and $O3$-planes.
In this case  the orientifold action is given by $\Omega \sigma (-1)^{F_L}$,
where $\sigma$ is a holomorphic, isometric involution acting
on the Calabi-Yau threefold ${\cal M}$.

The notion of poly-instantons \cite{Blumenhagen:2008kq,Blumenhagen:2008ji} means the correction
of an Euclidean D-brane instanton action  by other D-brane instantons.
The configuration we considered has two
instantons $a$ and $b$ with
proper zero modes to generate a  non-perturbative contribution
to the superpotential of the form
\eq{
W=A_a\, \text{exp}^{-S_a}+A_a A_b\, \text{exp}^{-S_a-S_b}+ ...\, ,
}
where $A_{a,b}$ are moduli dependent one-loop determinants and $S_{a,b}$
denote the classical $D$-brane instanton actions.

In Type IIB  orientifolds models, the
sufficient conditions for the  zero mode structures
for  poly-instanton corrections to the superpotential  have been
worked out  in \cite{Blumenhagen:2012kz}(for orientifold with $O5$- and $O9$- plane, see \cite{Blumenhagen:2008ji}).
Both instantons, $a$ and $b$, should be $O(1)$ instantons,
i.e.\ a single instanton placed in an orientifold invariant position with
an $O(1)$ projection which corresponds to an $SP-$type projection
for a corresponding  space-time filling $D7$-brane.
Instanton $a$ is an Euclidean $E3$ instanton wrapping a rigid divisor $E$ in
the Calabi-Yau threefold  with $H^{1,0}(E,{\cal O})=H^{2,0}(E,{\cal O})=0$ 
while instanton $b$ is  an Euclidean $E3$-brane  instanton
wrapping a divisor which admits a single complex Wilson line Goldstino,
i.e.\ a so-called  Wilson line divisor with equivariant
cohomology $H^{*,0}(W,{\cal O})=(1_+,1_+,0)$ under the involution $\sigma$.
In fact, the sufficient condition for a geometric configuration to support the
poly-instanton correction is that it precisely contains one
Wilson line modulino in $H^1_+(E,\cal O)$.

Some concrete Calabi-Yau threefolds both with and without $K3$ fibration structure were presented in \cite{Blumenhagen:2012kz} 
  which
featured all the requirements mentioned above.
For the present purpose, we focus on one particular model which
not only admits a Wilson line divisor $W$ but also rigid and
shrinkable del Pezzo divisors. Such geometries are particularly
interesting for studying moduli stabilization as they give rise to a
swiss-cheese type K\"ahler potential.
The Calabi-Yau threefold $\cal M$ is given by a hypersurface
in a toric variety with defining data,
\begin{table}[ht]
  \centering
  \begin{tabular}{c|cccccccc}
     & $x_1$  & $x_2$  & $x_3$  & $x_4$  & $x_5$ & $x_6$  & $x_7$ & $x_8$        \\
    \hline
    2  & -1 & 0 & 1 & 1 & 0 & 0 &  0 & 1  \\
    4  & -2 & 0 & 2 & 2 & 1 & 0 &  1 & 0  \\
    2  & -3 & 0 & 2 & 1 & 1 & 1 &  0 & 0  \\
    2  & 1 & 1 & 0 & 0 & 0 & 0 &  0 & 0  \\
  \end{tabular}
 \end{table}

\noindent
with Hodge numbers $(h^{21}, h^{11}) = (72, 4)$ and the corresponding Stanley-Reisner ideal
\begin{equation}
{\rm SR}=  \left\{x_1\,x_2,  x_4\,x_7,  x_5\,x_7,  x_1\,x_4\,x_8,
x_2\,x_5\,x_6,  x_3\,x_4\,x_8,  x_3\,x_5\,x_6,  x_3\,x_6\,x_8 \right\}\,.
\end{equation}
This geometry admits two inequivalent orientifold projections $\sigma: \{x_4
\leftrightarrow -x_4, x_7 \leftrightarrow -x_7\}$ with $h^{11}_-({\cal M})=0$
so that for the Wilson line divisor $W=D_8=\{x_8=0\}$ 
the Wilson line Goldstino is in $H_{+}^1(W,\cO)$. It was
checked that the $D3$- and $D7$-brane tadpoles can be canceled. For the
analysis in the following sections, we focus on the
involution $x_7 \leftrightarrow -x_7$.
The corresponding topological data of the relevant divisors are shown
in table \ref{tabledivsA}.
\begin{table}[ht]
  \centering
  \begin{tabular}{c|c|c}
    divisor & $(h^{00},h^{10},h^{20},h^{11})$  &  intersection curve    \\
    \hline \hline
      &    &  \\[-0.4cm]
    $D_7=dP_7$ & $(1_+,0,0,8_+)$ & $W: C_{g=1}$ \\
    $D_5$ & $(1_+,0,1_+,21_+)$ & $W: C_{g=1}$ \\
    \hline
    $D_8=W$ & $(1_+,1_+,0,2_+)$ & $D_5: C_{g=1}, \ \ D_7:  C_{g=1}$ \\
    $D_1=\IP^2$ & $(1_+,0,0,1_+)$ & $D_5: C_{g=0}$
  \end{tabular}
  \caption{\small Divisors and their equivariant cohomology under $x_7 \leftrightarrow -x_7$. The first two
    lines are $O7$-plane components  and the remaining two divisors
    can support $E3$ instantons. The $D_7$ divisor also supports
    gaugino condensation.}
  \label{tabledivsA}
\end{table}

\noindent
We also showed  that
 there are no extra vector-like zero modes on the intersection of $E3\cap D7$,
i.e.\  all sufficient conditions were satisfied for the divisor $W$
to generate a poly-instanton correction to the  non-perturbative superpotential
\eq{
   W&= A_1\, \exp\left( -2\pi T_1\right) + A_1\, A_8\, \exp\left(
       -2\pi T_1-2\pi T_8\right)+\\
&\phantom{aaaaaaaaaaa}A_7\, \exp\left( -a_7 T_7\right) + A_7\, A_8\, \exp\left(
       -a_7 T_7-2\pi T_8\right)+\ldots\; .
}

\noindent
Taking into account the K\"ahler cone constraints, the volume form for this
model could  be written in the strong swiss-cheese like form
\eq{
\label{volumeA}
{\cal V}&=\textstyle{ \frac{1}{9}}\Bigl(\frac{1}{\sqrt 2}
(\tau_1+3\tau_6+6\tau_7+3\tau_8)^{3/2}-\sqrt{2}\tau_1^{3/2}-3\tau_7^{3/2}-3(\tau_7+\tau_8)^{3/2}\Bigr)\,.
}
This volume form indicates that the large volume limit is given by
 $\tau_6 \rightarrow \infty$ while keeping the other shrinkable del Pezzo
four-cycles volumes $\tau_{1,7}$ and the Wilson line four-cycle volume $\tau_8$ small.
 In fact,  all the models studied in \cite{Blumenhagen:2012kz} shared a
similar strong swiss-cheese {\it like} volume form with the same intriguing
appearance of the Wilson line K\"ahler modulus.

\subsection{Moduli stabilization}
\label{sec_Moduli stabilization}

A generic orientifold compactification of Type IIB string theory
leads to an effective four-dimensional ${\cal N}=1$ supergravity theory \footnote{For a review on moduli stabilization and relevant compactification geometries, see \cite{Grana:2005jc,Blumenhagen:2006ci}.}.
In the closed string sector, the bosonic part of the 
massless chiral superfields arises from
the dilaton, the complex structure and  K\"ahler moduli and
the dimensional reduction of the NS-NS and R-R $p$-form fields.
The bosonic field content is given by
\eq{
\label{eq:N=1_coords}
 \tau&=C^{(0)}+ie^{-\phi} \, , \qquad  U^i = u^i + i v^i, \quad
   i=1\ldots h^{21}_+ \,, \\[0.1cm]
 {G}^a & = c^a - \tau {b}^a \,, \qquad a=1,\ldots, h^{11}_- \,, \\
 T_\alpha&=\frac{1}{2} \kappa_{\alpha\beta\gamma}t^\beta t^\gamma +
 i\left(\rho_\alpha -\kappa_{\alpha a b} \, {c^a b^b}\right) +\frac{i}{2}\,
 \tau \kappa_{\alpha ab} b^a b^b \quad \text{~and} \quad  \alpha=1,\ldots, h^{11}_+\,,
}
where $c^a$ and $b^a$ are defined as integrals of the
axionic $C^{(2)}$ and $B^{(2)}$ forms and $\rho_{\alpha}$ as
integrals of $C^{(4)}$ over a basis of four-cycles $D_\alpha$.
From now on, as in our aforementioned concrete example, we assume $h^{11}_-=0$.

The supergravity action is specified by the K\"ahler potential,
the holomorphic superpotential $W$ and the holomorphic gauge kinetic function.
The K\"{a}hler potential for the supergravity action is given as,
\begin{eqnarray}
\label{eq:K}
& & \hskip -1cm K = - \ln\biggl(-i(\tau-{\bar\tau})\biggr)
-\ln\left(-i\int_{\cal M}\Omega\wedge{\bar\Omega}\right)-2\ln\Bigl( {\cal V}(T_\alpha)\Bigr)\,,
\end{eqnarray}
where ${\cal V}={\frac 1 6} {\kappa}_{\alpha\beta\gamma} t^\alpha t^\beta t^\gamma$
is   the volume of the internal Calabi-Yau threefold. The general form of the superpotential $W$ is given as
\begin{equation}
\label{eq:W}
W = \int_{\cal M} G_3\wedge\Omega + \sum_{E} {A}_{E}(\tau, U^i) \,
e^{- a_E \gamma^\alpha\, T_{\alpha}} \,
\end{equation}
with the instantonic divisor given by $E=\sum  \gamma^\alpha D_\alpha$.
The first term is the Gukov-Vafa-Witten (GVW)  flux induced
superpotential \cite{Gukov:1999ya}  and the second one denotes  
the non-perturbative
correction coming from Euclidean $D3$-brane instantons ($a_E=2\pi$) and
gaugino condensation on $U(N)$ stacks of $D7$-branes ($a_E=2\pi/N$) \cite{Witten:1996bn}.
In terms of the  K\"{a}hler potential and the superpotential
the scalar potential is given by
\eq{
\label{eq:Vgen}
V = e^{K}\Biggl(\sum_{I,\,J}{K}^{I\bar{J}} {\cal D}_I W {\bar{\cal D}}_{\bar J } {\bar W} - 3 |W|^2 \Biggr)\,,
}
where the sum runs over all moduli.
As usual in the  LARGE volume scenario \cite{Balasubramanian:2005zx}, the complex structure moduli
and the axio-dilaton
are stabilized  at order $1/{\cal V}^2$ by $K$ and the GVW-superpotential respectively.
Since the stabilization of the K\"{a}hler moduli is by sub-leading
terms in the ${\cal V}^{-1}$ expansion, for this purpose
the complex structure moduli and the dilaton can be treated as
constants.

Let us proceed with the following ansatz for K\"{a}hler and superpotential\footnote{We consider the racetrack form of superpotential as it has been realized in \cite{Blumenhagen:2012ue} that the same is needed for having a (non-susy) minimum which could be trusted in the effective supergravity.} which is well motivated by our mathematical discussion in the previous subsection \ref{sec_Poly-instanton corrections}, 
\eq{
\label{eq:KW}
 K &= - 2 \, \ln {\cal Y} \,, \\[0.2cm]
W &= W_0 + A_s \, e^{-a_s T_s}+ A_s A_w \, e^{-a_s T_s - a_w T_w} \\
& \qquad \qquad- B_s \,
e^{-b_s T_s}- B_s B_w \, e^{-b_s T_s - b_w T_w} \, ,
}
where ${\cal Y}= {\cal V}(T_\alpha)+C_{\alpha^\prime}\;$ such that
\eq{
{\cal Y}
= \xi_b (T_b+\bar{T}_b)^{\frac{3}{2}}-\xi_s
(T_s+\bar{T}_s)^{\frac{3}{2}}-\xi_{sw}
\Bigl((T_s+\bar{T}_s) + (T_w+\bar{T}_w)\Bigr)^{\frac{3}{2}} +
C_{\alpha^\prime}\, .
}
Note that, for $h^{11}_{-} =0$, the ${\cal N} = 1$ K\"{a}hler
coordinates are simply given as $T_\alpha = \tau_\alpha + i \rho_\alpha$. Here, $C_{\alpha^\prime}$ denotes  the perturbative
${\alpha^\prime}^3$-correction  given as \cite{Becker:2002nn}
\eq{
C_{\alpha^\prime} = - \frac{\chi({\cal M}) \,
  ({{\tau}-\bar\tau})^{\frac{3}{2}} \zeta(3) }{4 (2 \pi)^3 \,
  ({2i})^{\frac{3}{2}}}
}
with $\chi({\cal M})$ being the Euler characteristic of the Calabi-Yau.
The large volume limit is defined by taking $\tau_b \rightarrow \infty$ while
keeping the other divisor volumes  small. 

In the large volume limit, the most dominant contributions to the
generic scalar potential ${\bf V}(\tau_b,\tau_s,\tau_w,\rho_s,\rho_w)$
are collected by three types of terms.
In the absence of poly-instanton effects, they  simplify to
\eq{
{\bf V}({\cal V},\tau_s,\rho_s) \simeq {\bf V}_{\alpha^\prime}({\cal V}) +
{\bf V}_{\rm np1} ({\cal V},\tau_s,\rho_s) + {\bf V}_{\rm np2}({\cal V},\tau_s,\rho_s)\,,
}
where
\bea
\label{eq:Vgen+racelvs}
& & \hskip -0.8cm {\bf V}_{\alpha^\prime}  = \frac{3 {\, {\cal
      C}_{\alpha^\prime}} \,|W_0|^2}{2 \, {\cal V}^3}\, ,\nonumber\\[0.1cm]
& & \hskip -0.8cm {\bf V}_{\rm np1} =\frac{4\, a_s \, {A_s}\, W_0\, e^{-\, a_s {\tau_s}} \,  {\tau_s} \cos(\, a_s {\rho_s})}{{\cal V}^2}
 -\frac{4\, b_s \, {B_s}\, W_0\,  e^{-\, b_s {\tau_s}} \,  {\tau_s} \cos (\, b_s {\rho_s})}{{\cal V}^2}\,,\\[0.1cm]
& & \hskip -0.8cm {\bf V}_{\rm np2} = \frac{2\sqrt2\, a_s^2 \, A_s^2 \, e^{-2 \, a_s {\tau_s}} \sqrt{{\tau_s}}}{3\,  \xi_s \, {\cal V}}-\frac{4 \sqrt2 \,
a_s \, b_s \, {A_s}\, {B_s}\, e^{-(\, a_s+\, b_s) {\tau_s}} \sqrt{{\tau_s}} \cos \bigl((\, a_s-\, b_s) {\rho_s}\bigr)}{3\, \xi_s \, {\cal V}}\nonumber\\
& &  \hskip 0.6cm +\frac{2 \sqrt2\, b_s^2 \, B_s^2\, e^{-2 \, b_s {\tau_s}}
   \sqrt{{\tau_s}}}{3\, \xi_s \, {\cal V}}\, .\nonumber
\eea
As expected, the potential \eqref{eq:Vgen+racelvs} does not
depend on the Wilson line  divisor volume modulus $\tau_w$.
Therefore, at this stage it  remains a flat-direction to be lifted
via sub-dominant poly-instanton effects.

The extremality conditions $\partial_{{\cal V}} {\bf V} =  \partial_{{\tau_s}} {\bf V}=\partial_{{\rho_s}} {\bf V}=0$ are  collectively given as
\eq{
\label{eq:Extremize}
W_0 &= \frac{\ov{\cal V} \,(b_s \ov\lambda_2 -a_s \ov\lambda_1) \, \Bigl[b_s \ov\lambda_2 (-1 + 4 b_s \ov\tau_s)
-a_s \ov\lambda_1 (-1 + 4 a_s \ov \tau_s)\Bigr]}{6 \sqrt2 \, \xi_s \, \sqrt{\ov\tau_s} \Bigl[b_s \ov\lambda_2 (-1 + b_s \ov\tau_s)-a_s \ov\lambda_1 (-1 + a_s \ov\tau_s)\Bigr]}\,,\\[0.2cm]
{\cal C}_{\alpha^\prime} &= \frac{32 \sqrt{2} \, \xi_s \, \ov\tau_s^{\frac{5}{2}} (b_s^2\, \ov\lambda_2 -
a_s^2\, \ov\lambda_1) \, \Bigl[b_s \ov\lambda_2 (-1 + b_s \ov\tau_s)-a_s \ov\lambda_1 (-1 + a_s \ov\tau_s)\Bigr]}{\Bigl[ a_s \ov\lambda_1(-1 + 4 a_s \ov\tau_s)
- b_s \ov\lambda_2(-1 + 4 b_s \ov\tau_s) \Bigr]^2} \, ,\\[0.2cm]
 a_s\, \ov\rho_s &= N \pi \quad   {\rm with} \quad
\ov\lambda_1 = A_s\, e^{-a_s \ov\tau_s} \quad \text{~and~} \ \  \ov\lambda_2 = B_s\, e^{-b_s
  \ov\tau_s}\; .
}
One finds that $\tau_s$ gets stabilized in terms of
$C_{\alpha^\prime}$ as $\ov\tau_s \sim ({C_{\alpha^\prime}})^{\frac{2}{3}}$
and then ${\cal V}$ gets stabilized via an exponential term $\exp(a_s\tau_s)$
(encoded in $\lambda_i$'s) so that  the overall volume of the
Calabi-Yau threefold is exponentially large. 

Now, in addition to the leading order standard racetrack terms
\eqref{eq:Vgen+racelvs}, the generic scalar potential also involves sub-dominant contributions,  which are further suppressed by
powers of $\exp(-a_w \tau_w)$.
Collecting these sub-leading terms, the effective potential for $\tau_w, \rho_w$ becomes
\eq{
{\bf V}({\cal V},\tau_s,\tau_w,\rho_s,\rho_w) = {\bf V}^{\rm LVS}({\cal V},\tau_s,\rho_s)+
 {\bf V}(\tau_w,\rho_w) \,,
}
where, ${\bf V}^{\rm LVS}({\cal V},\tau_s,\rho_s)$ is the racetrack version of the standard large volume potential which stabilizes the K\"ahler  moduli $\tau_b$ (or ${\cal V}$) and $\tau_s$  at order ${\cal V}^{-3}$ and 
\eq{
 {\bf V}(\tau_w,\rho_w) &\sim \frac{4 W_0}{{\cal V}^2}\Bigl[ \lambda_1 \,
{A_w}\, e^{-\, {a_w} {\tau_w}} \, (a_s \tau_s + a_w \tau_w)\, \cos( {a_w}
{\rho_w}) \\
&\phantom{aaaaaaaaaaaaaa} -\lambda_2 \, {B_w}\, e^{- {b_w} {\tau_w}} \, (b_s {\tau_s} + b_w \tau_w)\, \cos({b_w} {\rho_w})\Bigr]\\
&+\frac{4\sqrt 2\, (b_s {\lambda}_2 -a_s {\lambda}_1) \, \sqrt{\tau_s}}{3\,\xi_s \,{\cal V}}
\Bigl[ \lambda_2 (b_s -b_w) B_w \, e^{- b_w \tau_w}
  \cos (b_w \rho_w) \\
 & \phantom{aaaaaaaaaaaaaaaaaaaaa} -\lambda_1 (a_s -a_w) A_w \, e^{- a_w \tau_w} \, \cos(a_w \rho_w) \Bigr]}.
After stabilizing the heavier moduli ${\cal V}, \tau_s$, and $ \rho_s$-axion at their respective minimum and
using $a_w = b_w$ along with eliminating $W_0$ via the first relation
in eq.\eqref{eq:Extremize},
the above effective scalar potential can be written as
\bea
\label{eq:Vtau8race}
& & {\bf V}(\tau_w, \rho_w) = V_0 + e^{-a_w \tau_w}\left(\mu_1 + \mu_2 \, \tau_w \right) \, \cos(a_w \rho_w)\, .
\eea
Here $V_0, \mu_1, \mu_2$ are constants depending on the stabilized values of the heavier moduli as
\eq{
\label{eq:Vtau8}
\mu_1 &= \mu_0 \biggl[4 \ov\tau_s \bigl( (a_s -a_w) A_w \ov\lambda_1 -(b_s -a_w) B_w \ov\lambda_2\bigr)\\
 &+\frac{\ov\tau_s\, (b_s B_w \ov\lambda_2-a_s A_w \ov\lambda_1)
\,\bigl(a_s \ov\lambda_1 (-1+4a_s \ov\tau_s)-b_s \ov\lambda_2 (-1+4b_s
  \ov\tau_s)\bigr)
}{a_s \ov\lambda_1 (-1+a_s \ov\tau_s)-b_s \ov\lambda_2 (-1+b_s \ov\tau_s)}
\biggr]\,,\\[0.2cm]
\mu_2 &= \mu_0 a_w \left[\frac{\bigl( B_w \ov\lambda_2- A_w \ov\lambda_1\bigr) \bigl(a_s \ov\lambda_1 (-1+4a_s \ov\tau_s)
-b_s \ov\lambda_2 (-1+4b_s \ov\tau_s)\bigr)}{a_s \ov\lambda_1 (-1+a_s
    \ov\tau_s)-b_s \ov\lambda_2 (-1+b_s \ov\tau_s)}\right]}
with
\eq{
\mu_0 = \frac{\sqrt2 (\, a_s \ov\lambda_1-b_s \ov\lambda_2)}{3 \, \xi_s
  \,\ov{\cal V} \, \sqrt{\ov\tau_s}}\, .}
After stabilizing the $\rho_w$-axion at its minimum $\ov \rho_w=0$, it has been shown (in \cite{Blumenhagen:2012ue}) that the divisor volume modulus $\tau_w$ corresponding to the Wilson divisor $D_w$ gets stabilized to 
\bea
\label{eq:Tau8race}
& & a_w \ov\tau_w = 1-a_w\frac{\mu_1}{\mu_2}\, ,
\eea
where $\f{\mu_1}{\mu_2}<0$ is required for stabilizing $\tau_w$ inside the K\"ahler cone.
A couple of benchmark models have been constructed for a set of sampling parameters displacing the K\"ahler modulus $\tau_w$ away from its minimum, investigations of inflationary behavior have been made in a single field approximation \cite{Blumenhagen:2012ue}. In this article, we generalize the inflationary model via including the dynamics of corresponding $\rho_w$ axion  and discuss various subsequent cosmological implications.

\section{Two-Field Poly-Instanton Inflation}
\label{sec_Inflationary Cosmology}

In all the remaining sections, we will be assuming that the heavier moduli are stabilized at their respective minimum position and we consider the effective potential for the two lighter fields ($\tau_w$ and $ \rho_w$)  given by \eqref{eq:Vtau8race}.
Furthermore, we assume that a suitable uplifting of the $AdS$ minimum
to a $dS$ minimum can be processed via an appropriate mechanism (e.g.~by the introduction of anti-D3 brane \cite{Kachru:2003aw}\footnote{For uplifting process using anti-D3 brane, there has been some sensitive issues as mentioned in \cite{Blaback:2012nf,Bena:2012vz,Bena:2012bk}.} or other mechanism like using  dilaton-dependent term in superpotential \cite{Cicoli:2012fh}). Thus, the resulting inflationary potential
looks like
\bea
\label{effective_potential}
& & \hskip-1.5cm V_{\rm inf}(\tau_w, \rho_w) = V_{\rm up} + V_0 + e^{-a_w \tau_w}\left(\mu_1 + \mu_2 \, \tau_w \right) \, \cos(a_w \rho_w)\,
\eea
Now, we state the effective potential studied before in roulette inflation \cite{Bond:2006nc} which after stabilizing all but one K\"ahler moduli effectively looks as,
\bea
\label{roulette}
V_{roulette}(\tau_n, \rho_n) = V_{\rm up} + V_0 +\frac{\sqrt{\tau_n} \, e^{-2 a_n \tau_n}}{\ov{\cal V}} + \frac{\tau_n \, e^{-a_n \tau_n} \, \cos(a_n \rho_n)}{{\ov{\cal V}}^2}
\eea
where $\tau_n$ is stabilized at $\ov\tau_n \sim {\rm ln}{\ov{\cal V}}$. 
Let us mention some important differences between the aforementioned potentials.
First, the geometric origin of these two potential are different as we have explained in the previous sections. The former one comes
from a Wilson divisor volume modulus while the later one comes from a blow-up volume modulus.
Second, these moduli are stabilized by corrections which scales differently in terms of CY volume. Moreover, the stabilized values at the respective minimum have different volume scaling; $\ov\tau_n$ is stabilized at order ${\rm ln} \ov{\cal V}$ while $\tau_w$ is stabilized by eq.(\ref{eq:Tau8race}). Here, the $\ov{\cal V}$ dependence, which could have appeared via $\ov\tau_s$, gets {\it effectively} canceled.

\begin{figure}[ht!]
\centering
\includegraphics[scale=0.6]{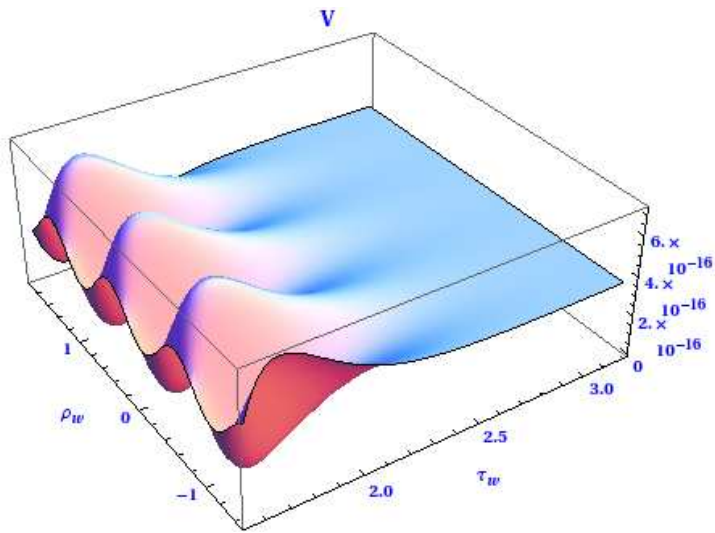}
\caption{The effective potential as a function of the moduli $\tau_w$ and $\rho_w$.
 }\label{potential}
\end{figure}

The uplifting term $V_{\rm up}$ in eq.(\ref{effective_potential}) needs to be such that the
uplifted scalar potential acquires a small positive value (to be matched with
the cosmological constant) when all the moduli sit at their respective
minimum. This potential has the following set of critical points 
\bea
& & (i).   \hskip1.0cm \ov\tau_w = \frac{\mu_2 - a_w \, \mu_1}{a_w \, \mu_2} , \, \, \,  a_w \ov\rho_w =2 p \, \pi \\
& & (ii).   \hskip0.88cm \ov\tau_w = \frac{\mu_2 - a_w \, \mu_1}{a_w \, \mu_2} , \, \, \,  a_w \ov\rho_w = (2 p +1 ) \pi \nonumber
\eea
where $p\in \mathbb Z$. Moreover, in order to trust the effective field theory we need $\f{\mu_1}{\mu_2}<0$. In order to ensure the minimum, one has to consider the Hessian $V_{ab}$ (evaluated at these two sets of critical points) which is given as 
\bea
\label{eq:Hess2}
\hskip-1cm V_{ab}= \left(
\begin{array}{ll}
\mp a_w\, \mu_2 \, e^{-1+\frac{a_w \, \mu_1}{\mu_2}}& \, \, \, \, \, \, \, \, \, \, \, \, \, 0   \\
\, \, \, \, \, \, \, \, \, \, \, \, \, \, \, \, 0& \mp a_w\, \mu_2 \, e^{-1+\frac{a_w \, \mu_1}{\mu_2}} 
\end{array}
\right)\,,
\eea
where $\mp$ sign corresponds to critical points $(i)$ and $(ii)$ respectively. Hence depending on whether $\{\mu_1 <0, \mu_2 > 0\}$ or $\{\mu_1 >0, \mu_2 < 0\}$, we find that one set of critical point corresponds to minima while the other corresponds to maxima. From now on, we fix our notation with a sampling of parameters such that $\{\mu_1 >0, \mu_2 < 0\}$ (as in \cite{Blumenhagen:2012ue}) and performing the redefinitions $\tau_w = \phi_1, \, \rho_w = \phi_2$, the uplifted scalar potential takes the form 
\bea
\label{eq:Vinf}
& & \hskip-1.3cm V_{\rm inf} (\phi_1, \phi_2)=\left(\frac{g_s}{8 \pi}\right)\, e^{K_{\rm CS}} \, \left[-\frac{\mu_2 \, e^{-1+\frac{a_w \, \mu_1}{\mu_2}}}{a_w} + e^{-a_w \phi_1}\left(\mu_1 + \mu_2 \, \phi_1 \right) \, \cos(a_w \phi_2)\,\right].
\eea
Here, a proper normalization factor $\left(\frac{g_s}{8 \pi}\right)\, e^{K_{\rm CS}}$ has been included \cite{Conlon:2005jm}, where $K_{\rm CS}$ denotes the K\"ahler potential for the complex structure moduli. We assume that $e^{K_{\rm CS}}\sim{\cal O}(1)$. Furthermore, we set the numerical parameters for moduli stabilization similar to the ones chosen in one of the benchmark models (in \cite{Blumenhagen:2012ue}). The parameters which would be directly relevant for further computations in this article are,
\bea
\label{sampling}
 & & \mu_1 = 2.9\times10^{-8}, \, \mu_2 = -1.9\times 10^{-8}, \, a_w = 2 \pi, g_s = 0.12\\
& &  \hskip 2cm \ov{\cal V} = 905, \ov\tau_s = 5.7, \xi_{sw} = 1/(6 \sqrt2) \nonumber
\eea
The `effective' non-flat moduli space metric ${\cal G}_{ab}$ relevant for inflaton dynamics can be computed as 
\bea
\label{eq:Gab}
 {\cal G}_{ab}= \left(
\begin{array}{ll}
\frac{3 \xi_{sw}}{2\,\sqrt2 \, {\cal V}\, \sqrt{{\tau_s}+{\phi_1}}}&  \, \, \, \, \, \, \, \, \, \, \, \,  0   \\
  \, \, \, \, \, \, \, \, \, \, \, \, \, 0 & \frac{3 \xi_{sw}}{2\,\sqrt2 \, {\cal V}\, \sqrt{{\tau_s}+{\phi_1}}} 
\end{array}
\right)\,
\eea
Utilizing the large volume limit, the same is written from the K\"ahler metric component $K_{T_w {\ov T}_w}$ expressed in the real moduli basis $\{\phi_1, \phi_2\}$. The Christoffel connections are defined as $\Gamma^a_{\,\, bc}= {\cal G}^{ad}\Gamma_{bc,d}= \frac{1}{2}{\cal G}^{ad}\left(\partial_b {\cal G}_{cd} +\partial_c {\cal G}_{bd} -\partial_d {\cal G}_{bc} \right)$ and the only non-zero connection components are 
$$\Gamma_{\,\, 11}^1=-\frac{1}{4(\ov\tau_3 + \phi_1)} = \Gamma_{\, \, 12}^2 = \Gamma_{\, \, 21}^2 , \, \, \Gamma_{\, \, 22}^1 = \frac{1}{4(\ov\tau_3 + \phi_1)}.$$
Furthermore, the non-zero components of the Riemann tensor, defined as $R^a_{\,\,\, bcd} = \partial_{c} \Gamma^{a}_{\, \, db}-\partial_{d} \Gamma^{a}_{\, \,cb }+\Gamma^{a}_{\, \, ck} \Gamma^{k}_{\, \,db}-\Gamma^{a}_{\, \, dk} \Gamma^{k}_{\, \,cb }$, are
$$R_{\,\, 212}^1=-\frac{1}{2(\ov\tau_3 + \phi_1)^2} = R_{\, \, 121}^2 , \, \, R_{\, \, 221}^1 = \frac{1}{2(\ov\tau_3 + \phi_1)^2} = R_{\, \, 112}^2.$$
Under the sampling (\ref{sampling}), the effective two-field inflationary potential from eq.(\ref{effective_potential}) is shown in Fig.\ref{potential}.

\section{Evolution of Trajectories}
\label{sec_Evolution}

In this section, we investigate the field evolutions and look for the possible inflationary trajectories which could reproduce 50 (or more) e-foldings. For a given initial condition, we numerically trace the entire inflationary trajectories including beyond slow-roll region. 
Using the background $N$ e-folding number as the time coordinate, i.e. $dN = H dt$, the Einstein-Friedmann equations are obtained as
\begin{subequations}
\bea
\label{Friedmann}
\f{d^2}{dN^2}\phi^a+{\Gamma^a}_{bc}\f{d\phi^b}{dN}\f{d\phi^c}{dN}+\left(3+\f{1}{H}\f{dH}{dN}\right)\f{d\phi^a}{dN}+\f{{\cal G}^{ab} \partial_b V}{H^2}=0 ,
\eea
\bea
\label{constran}
H^2=\f{1}{3}\left(V(\phi^a)+\f{1}{2}H^2 \, {\cal G}_{ab} \f{d\phi^a}{dN}\f{d\phi^b}{dN} \right).
\eea
\end{subequations}
Using expressions (\ref{Friedmann}) and (\ref{constran}), one can derive another useful expression for variation of Hubble rate in terms of e-folding, 
\bea
\label{third}
\f{1}{H}\f{dH}{dN}=\f{V}{H^2}-3 .
\eea
For numerical convenience, we solve these equations in the time basis $t$ and then change the result back to the basis $N$ e-folding. As introduced in \cite{Yokoyama:2008by}, we will  follow
the field redefinitions given as\footnote{The use of this notation would be more clear in the next section regarding computations of non-linearity parameters. Further, we will be using a combined indexing ${\cal A}$ such that any object ${\cal O}^{\cal A}$ has two components given as ${\cal O}^{\cal A}\equiv \{{\cal O}^a_1, {\cal O}^a_2\}$.}
\bea
\label{eq:redef}
& &  \varphi^{a}_1\equiv\phi^a, \,\,\,\,\,\,\  \varphi^{a}_2\equiv \dot{\phi}^a = \left(\f{d \phi^a}{d t}\right), \,\,\,\,\ {\rm where}\,\, a=1,\, 2.
\eea
which translates the second-order background equations of motions eq.(\ref{Friedmann}) into two first-order ODEs as follows
\bea\label{EOM}
& & F^a_1 \equiv \frac{d\varphi^a_1}{dN}=\left(\frac{d \phi^a}{dN}\right)= \f{\varphi^a_2}{H},\nonumber\\
& & F^a_2 \equiv \frac{D\varphi^a_2}{dN}= -3 \varphi^a_2 - {\cal G}^{ab} \f{V_b}{H}.
\eea
where $D$ is the covariant derivative defined as $D \varphi^a_2 = d \varphi^a_2 + {\Gamma^a}_{bc} \varphi^b_2  d\varphi^c_1$, subject to the constraints 
$H^2=\f{1}{3}\left(V+\f{1}{2}\, {\cal G}_{ab} \varphi^a_2 \varphi^b_2 \right)$. Then eq.(\ref{third}) will be simplified as $\f{dH}{dt}= -\f{1}{2}\, {\cal G}_{ab} \varphi^a_2 \varphi^b_2 $. Now, in the context of studying inflationary aspects, one has to look at the sufficient conditions for realizing slow-roll inflation which are encoded in the so-called slow-roll parameters. For multi-field inflationary process with inflatons moving in a non-flat background, these slow-roll parameters are given as,
\bea
 \epsilon \equiv -\f{1}{H^2}\f{dH}{dt}, \,\,\,\,\,\,\,\, \eta \equiv \f{1}{\epsilon H}\f{d \epsilon}{d t}.
\eea

\begin{figure}[ht!]
\centering
\includegraphics[scale=1]{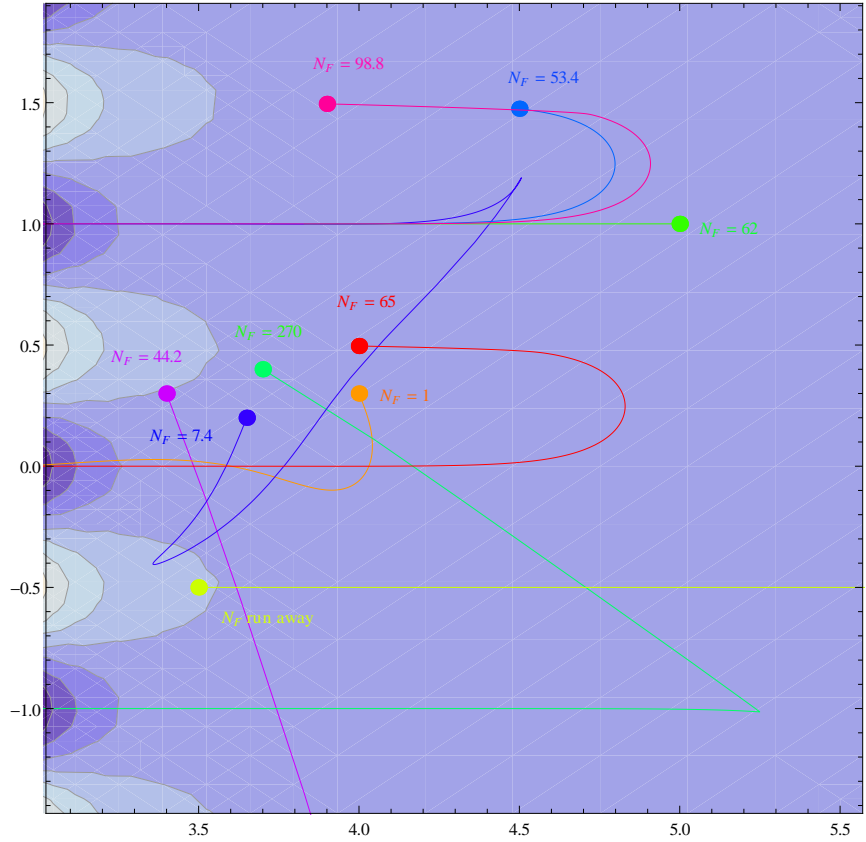}
\caption{The full inflationary trajectories for various initial conditions where the value of $N$ e-folding at the end of inflation is labeled on each of these trajectories.
 Various minima in dark blue are separated by maxima in light blue shade.
 }\label{trajectory}
\end{figure}

\begin{figure}
\centering
 \includegraphics[width=0.4\textwidth]{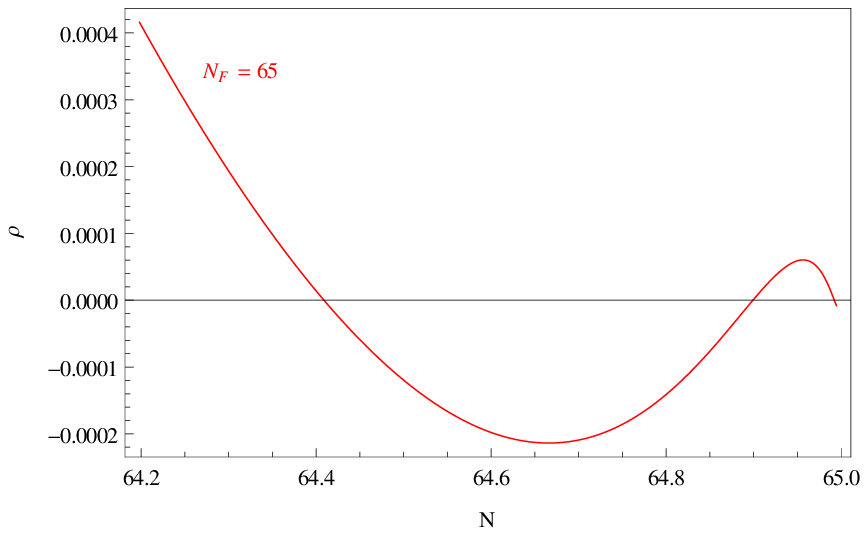}
 \includegraphics[width=0.37\textwidth]{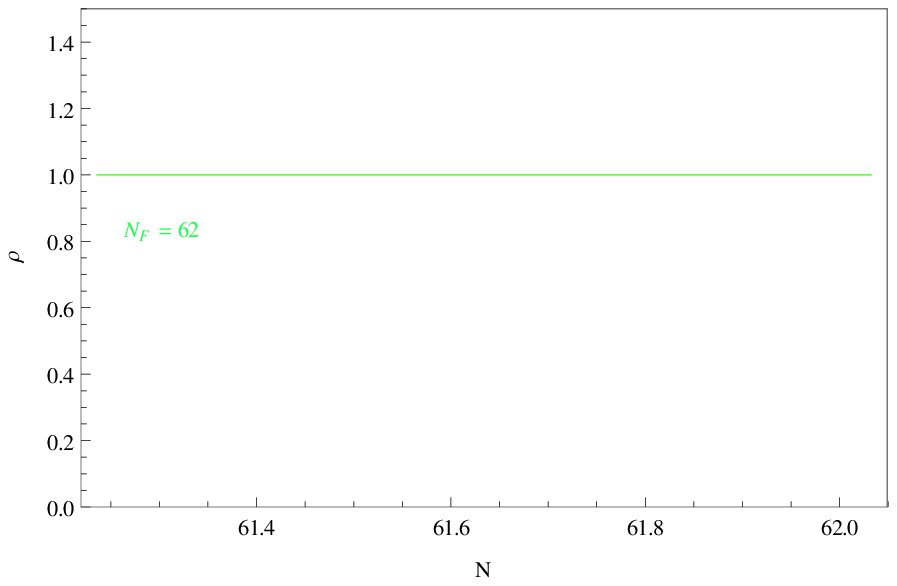}
 \\
 \includegraphics[width=0.4\textwidth]{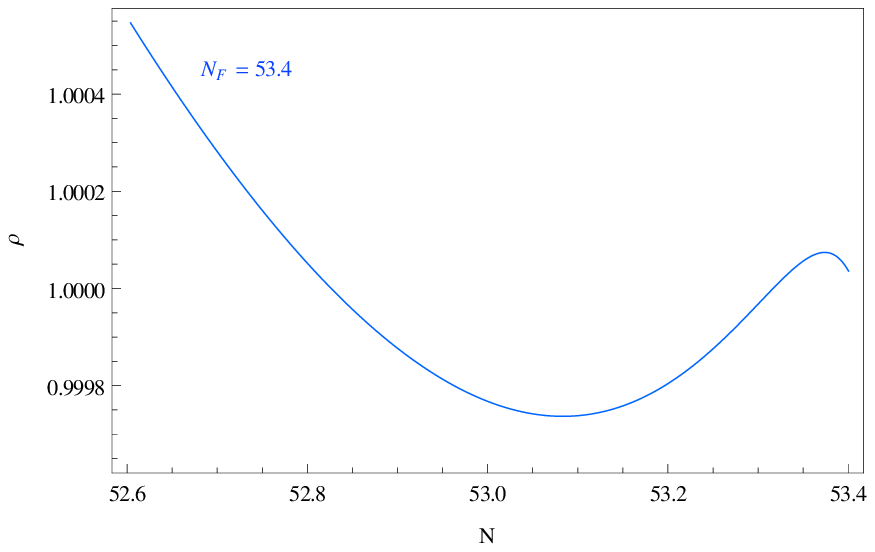}
 \includegraphics[width=0.4\textwidth]{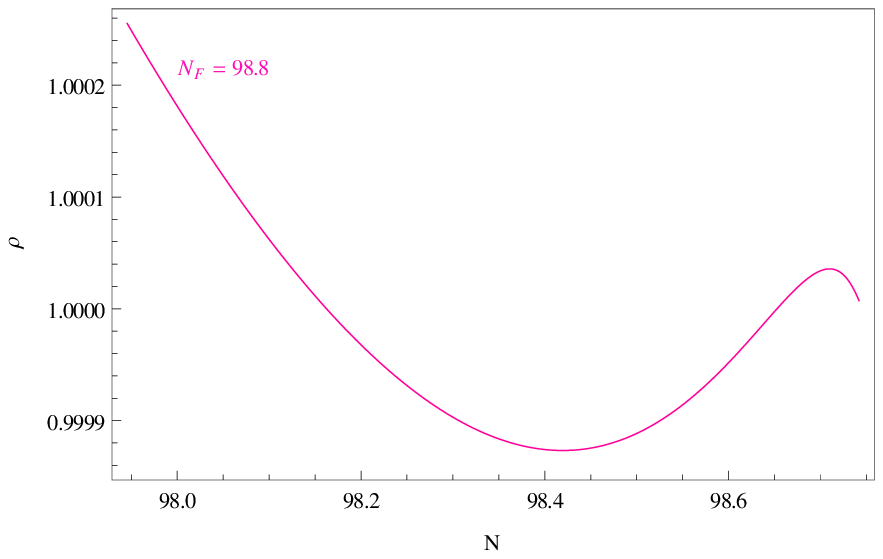}
  \caption{Oscillation of axion-field $\rho_w$ during late period in the beyond slow-roll regime. The $N$ e-folding when the $\rho_w$ crosses its  minimum is denoted as $N^{\dagger}= \{ 64.67,\,{\rm NULL},\,53.08,\,98.42 \}$. At the time $N^\dagger$, the divisor volume mode is still quite far from its minimum and thus both fields are dynamical and it is already in beyond slow-roll regime. Notice that the second trajectory corresponds to a single-field inflation. }
  \label{oscillate}
\end{figure}

Now, we can solve the background field equations (\ref{EOM}) to get the full trajectories under different initial conditions. We choose  $\phi^a(0)=\phi^a_0 \, \, {\rm and} \, \, \f{d\phi^a}{dt}\f{d\phi_a}{dt}|_{t=0}=0;\, \, {\rm for} \, \,  a \in \{1,2\}$  
as  a set of initial conditions and trace the corresponding trajectories up to the end of inflation (see Table \ref{initial}).
Figure \ref{trajectory} shows the complex evolution of trajectories for some samples of general starting conditions where the value of $N$ e-folding at the end of inflation is labeled on each of these trajectories. Note, although it looks like a single field inflation in the late period of trajectory, the axion field in fact oscillates (Figure \ref{oscillate}) during that time  before settling into its minimum. The fact that both of the fields in each trajectories keep evolving up to the end of inflation might be relevant for generating large Non-Gaussianity. This indicates that isocurvature modes are not completely finished even though it looks as a single field motion in the final stage of the trajectories. The $N$ e-folding when the axion $\rho_w$ first crosses its  minimum is denoted as $N^{\dagger}$. One has to note that at this time $N^\dagger$, the Wilson divisor volume mode is still quite far from its respective minimum. However, it rolls quickly enough to reach there. All of these points $N^{\dagger}
$ 
are already in 
the beyond slow-roll region as the slow-roll parameter $\eta>1 $.

\begin{figure}[ht!]
\centering
\includegraphics[scale=0.7]{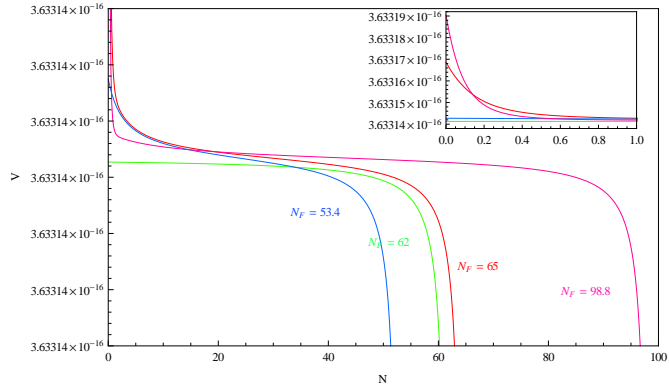}
\caption{The potential $V$ as a function of e-folding $N$, where the value of $N$ e-folding at the end of inflation  is labeled on each of these trajectories.}
\label{Vn}
\end{figure}

The various inflationary trajectories shown in Figure.\ref{trajectory} can be classified in the following categories
\begin{table}[ht]
  \centering
  \begin{tabular}{|c||c|c||c|}
  \hline
    Class & $\tau_w$  & $\rho_w$  & $N_F$     \\
    \hline \hline
     a  &  5 & 1 & 62   \\
  \hline 
   & 4 & 0.3 & 1 \\  
  b   & 4.5 & 1.475 & 53.4   \\ 
   & 4 & 0.496 & 65  \\  
     & 3.9 & 1.495 & 98.8  \\
     \hline
     c  & 3.5 & -0.5 & -  \\
    \hline
    & 3.65 & 0.2 & 7.4  \\  
  d & 3.4 & 0.3 & 44.2 \\
      & 3.7 & 0.4 & 270 \\
    \hline
  \end{tabular}
  \caption{Initial conditions for these trajectories shown in Figure.\ref{trajectory}.}
  \label{initial}
 \end{table}
\begin{enumerate}[(a)]
\item{If the axion initial condition is such that the axion is minimized at its respective minimum, then two-field inflationary process reduces to its single field analogue which has been studied in \cite{Blumenhagen:2012ue}. These are stable trajectories and are attracted towards the respective valley in a straight line like the trajectory in Figure.\ref{trajectory} with $N_F=62$. These can produce the required number of e-foldings if the Wilson divisor volume mode is displaced significantly away from the minimum.  }
\item{If the axion initial condition is a little bit away from the minimum, the trajectories rolls to the nearest valley and trace towards the respective minimum like those trajectories in Figure.\ref{trajectory} with $N_F=1, \, 53.4, \, 65, \, 98.8$.}
\item{If the axion initial condition starts with its value at the maximum, this results in an unstable trajectory directed straightly outwards from the respective attractor point showing a run-away behavior like the yellow trajectory in Figure.\ref{trajectory}.}
\item{The trajectories starts from  axion initial conditions being closer (but not exactly equal) to some maximum value as well as the initial values for the divisor volume mode being not very far from its respective minimum, one observes that inflationary trajectories cross several axion-ridges before getting attracted into a valley. This can be understood from the fact that this class of initial conditions is such that the initial potential energy is just a little higher to begin with (as shown in Figure.\ref{Vn} ) and the $N$ e-folding increase very slow at the beginning of these trajectories, see Figure.\ref{trajectory} with $N_F=7.4,\, 44.2, \, 270$.}
\end{enumerate}
\begin{figure}[ht!]
\centering
  \includegraphics[width=0.55\textwidth]{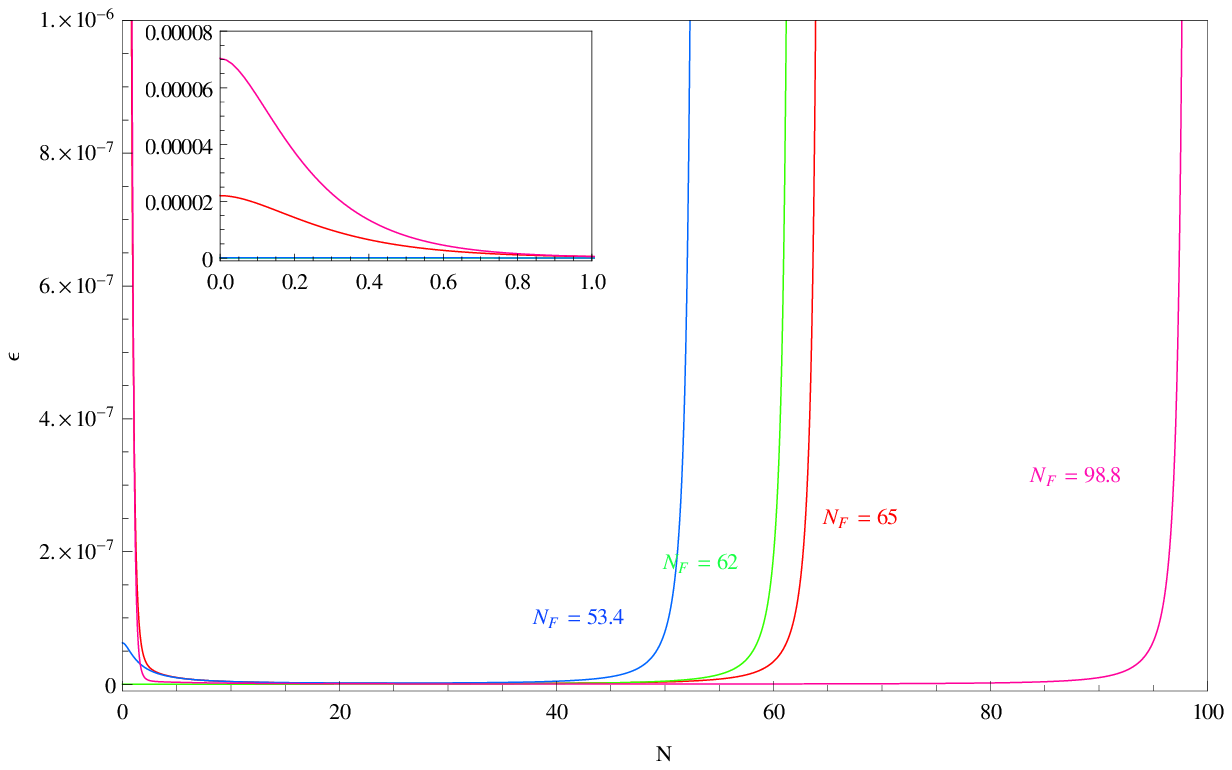}
  \includegraphics[width=0.53\textwidth]{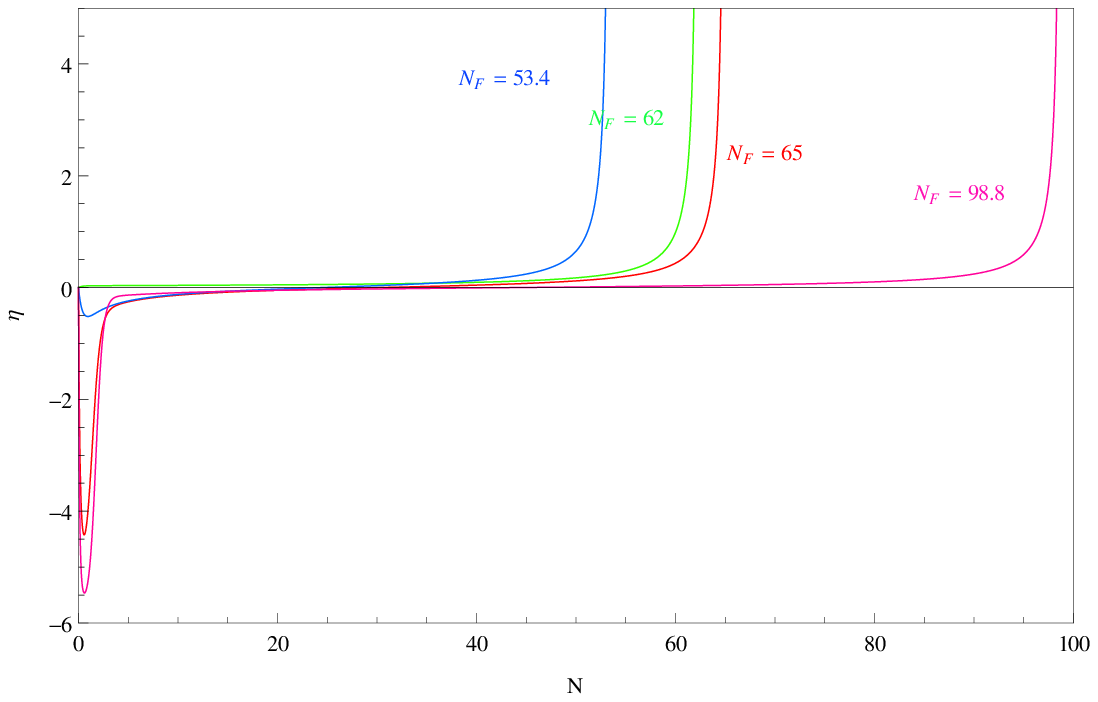}
\includegraphics[width=0.53\textwidth]{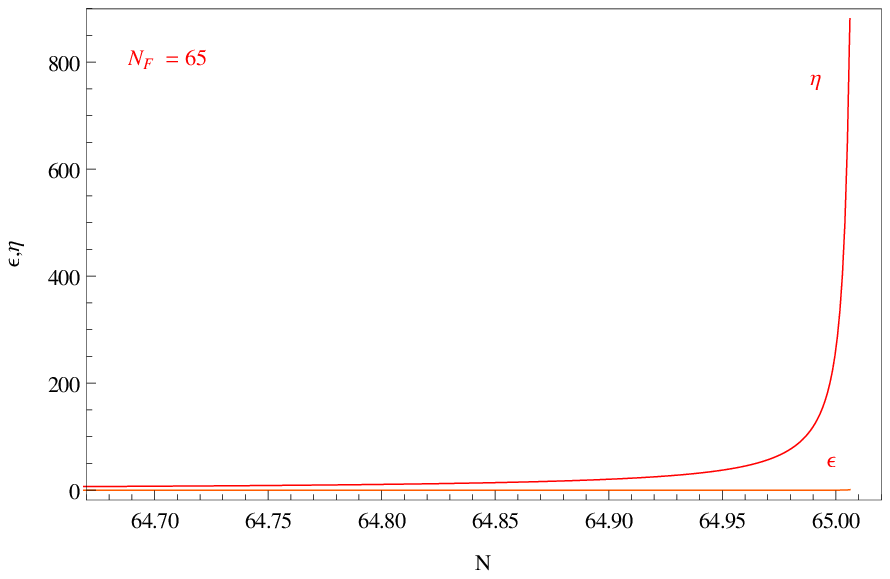}
  \caption{Slow-roll parameter $\epsilon$ and $\eta$ as a function of e-folding $N$, where the value of $N$ e-folding at the end of inflation is labeled on each of these trajectories. The enhancement of the $\eta$ parameter towards the end of inflation is illustrated for one trajectory corresponding to $N_F = 65$ in the last plot.}
  \label{fig:epsiloneta}
\end{figure}
\begin{figure}[ht!]
\centering
  \includegraphics[width=0.55\textwidth]{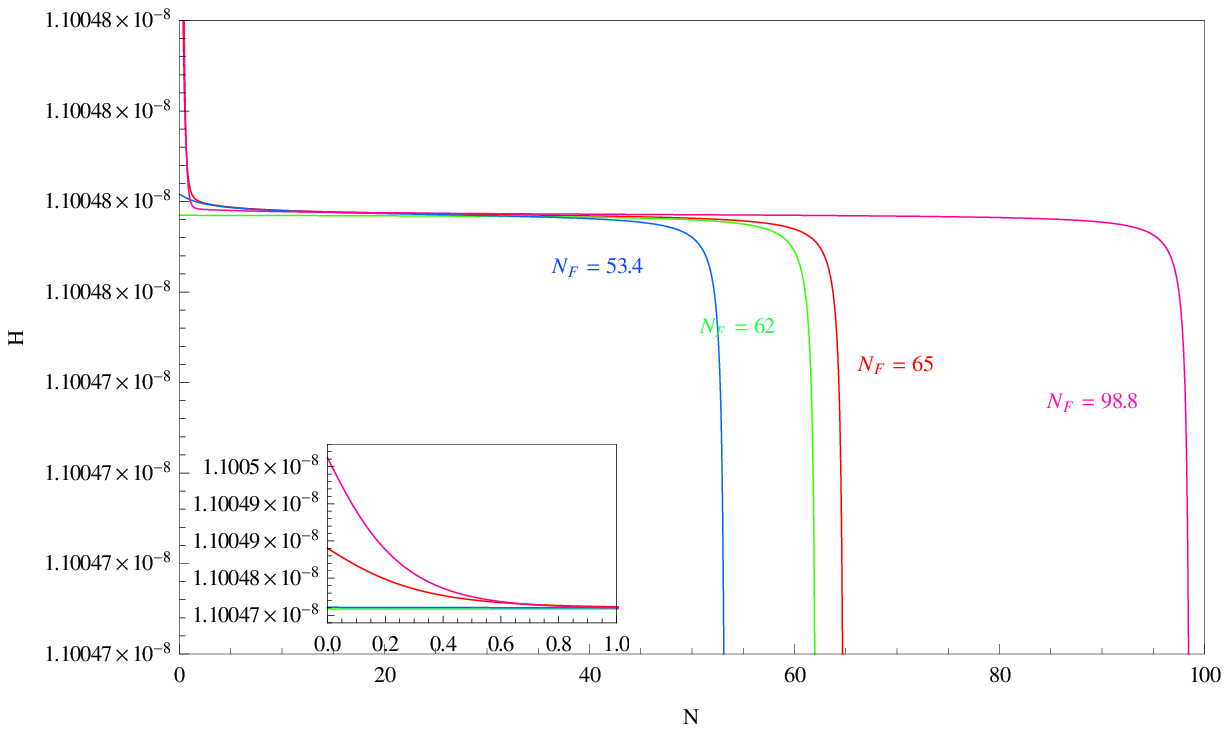}
  \includegraphics[width=0.55\textwidth]{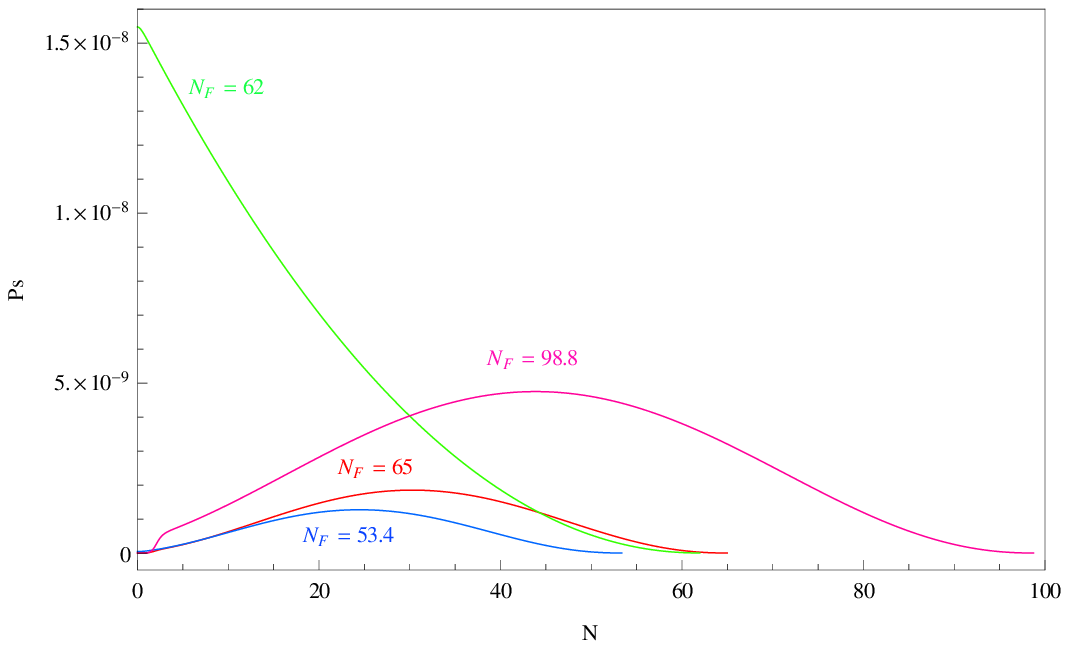}
  \caption{The Hubble rate $H$ and the power spectrum $P_s$ as a function of e-folding $N$, where the value of $N$ e-folding at the end of inflation is labeled on each of these trajectories. }
  \label{fig:HPs}
\end{figure}

For most of these trajectories except the single-field one, they momentarily undergo some sort of quick-roll region before staring the slow-roll (see Figure \ref{Vn}, \ref{fig:epsiloneta} and \ref{fig:HPs}). 
The slow-roll region ends when any of the slow roll conditions $\epsilon \ll 1, \, \eta \ll 1$ is violated. The evolution of the slow-roll parameter $\epsilon$ and $\eta$ depending on the variation of e-folding for four of these trajectories are shown in Figure \ref{fig:epsiloneta} from which one can see that these parameter changes dramatically at the end of inflation. 
Also, there is a region in field space where there is a {\it strong} violation of slow-roll condition via $\eta \gg 1$ before the end of inflation. This beyond slow-roll region can be interesting from the non-Gaussianities point of view as explained later. Furthermore, we can 
define the scalar perturbation power spectrum $P_s$ (in the slow-roll region) as $P_s = \f{H^2}{(2\pi)^2 \epsilon}$. Figure \ref{fig:HPs} shows the evolution of the Hubble rate $H$  and power spectrum $P_s$ with e-folding variations for the corresponding trajectories.
It shows that the Hubble rate $H$ is almost constant (at $10^{-8} M_p$) during entire inflationary process including the beyond slow-roll region also. This indicates a high scale inflation as $M_{\rm inf} \sim V_{\rm inf}^{\f{1}{4}} \sim H^{\f{1}{2}} \sim 10^{14} $GeV. The power spectrum for scalar perturbations $P_s$ reaches a value of order $10^{-9}$ at the horizon exit. Also, within the slow-roll limit, the spectral index is found to be negligibly small. All of these results are consistent with the observational constraints today.

\section{Primordial Non-Gaussianities}
\label{sec_fNL}
The signatures of non-Gaussianities are encoded in a set of non-linearity parameters which are commonly denoted as $f_{NL}, \tau_{NL}$ and $g_{NL}$.  These are generically related to the n-point correlators of curvature perturbations; the 2-point correlators simply give rise to a Gaussian shaped power spectrum while the 3-point correlators are related to the bi-spectrum which encodes the non-Gaussianities via the non-linearity parameter $f_{NL}$. Similarly, the 4-point correlators give rise to a tri-spectrum via $\tau_{NL}$ and $g_{NL}$ parameters. These non-linearity parameters can be computed in the $\delta N$-formalism which relates the curvature perturbations to the difference between the number of e-foldings $\delta N$ of two constant time-hypersurfaces \cite{Maldacena:2002vr},
$$\zeta(t,x) \simeq \delta N = H \delta t$$
Following the redefinitions of the background field evolutions as made in the previous section, the perturbations of the scalar field on $N = {\rm constant}$ gauge can be expressed as,
$$\delta \varphi^{\cal A}(\lambda, N) \equiv \varphi^{\cal A}(\lambda + \delta\lambda, N)-\varphi^{\cal A}(\lambda, N)$$ where $\lambda$'s are $2 {n}-1$ integration constants (for an $n$- component scalar field) which, along with $N$, parametrizes the initial values of the fields \cite{Yokoyama:2007dw,Yokoyama:2008by}. The curvature perturbations at each spatial point of the field space are subsequently expressed in terms of variations of the number of e-foldings in various field directions,
\bea
& & \zeta(N_F, {\bf x}) = \sum\frac{1}{n !} N^*_{{\cal A}_1{\cal A}_2....{\cal A}_n} \, \delta\varphi^{{\cal A}_1}({\bf x}) \, \delta \varphi^{{\cal A}_2}({\bf x}) ......\delta \varphi^{{\cal A}_n}({\bf x}),\\
& & \hskip-0.3cm N^*_{{\cal A}_1{\cal A}_2....{\cal A}_n} \equiv \left(\frac{\partial^n N(N_F, {\varphi^{\cal A}})}{\partial\varphi^{{\cal A}_1}\partial\varphi^{{\cal A}_2}....\partial\varphi^{{\cal A}_n}}\right)_{at \, \, \, \varphi^{\cal A} = \varphi^{\cal A}_{(0)}(N_*)}\nonumber
\eea
where $\varphi^{\cal A}_{(0)}$ corresponds to an unperturbed trajectory and $N_F$ corresponds to a final time-hypersurface of uniform energy density. 
The non-linearity parameters $f_{NL}$, $\tau_{NL}$ and $g_{NL}$ are generically defined as,
\bea
\label{eq:fNLgNLtauNLgen}
& &  f_{NL} = \frac{5}{6} \, \frac{ N^{\cal A} \,  N^{\cal B} \, N_{{\cal A}{\cal B}}}{ (N^{\cal D} \,  N_{\cal D})^2}, \nonumber\\
& & \tau_{NL} = \, \frac{ N^{\cal A} \,  N_{{\cal A}{\cal B}} \, N^{{\cal B}{\cal C}} \, N_{\cal C}}{(N^{\cal D} \,  N_{\cal D})^3}, \\
& &  g_{NL} = \frac{25}{54} \, \frac{ N^{\cal A} \,  N^{\cal B} \, N^{\cal C} \, N_{{\cal A}{\cal B}{\cal C}}}{ (N^{\cal D} \,  N_{\cal D})^3}\nonumber
\eea
where the field variations of $N$ are defined as $N_{\cal A}=\partial_{\cal A}N, N_{{\cal A}{\cal B}} = \partial_{{\cal A}{\cal B}} N$ and $N^{\cal A} = {\cal G}^{{\cal A}{\cal B}} N_{\cal B}$. Now, the main task in computing the non-linearity parameters is to find out the fields variations of the number of e-foldings and for the same we would follow the strategy developed in \cite{Yokoyama:2007dw,Yokoyama:2008by} using $\delta N$ formalism. It is important to mention that this approach is valid in the beyond slow-roll regime as well. So one can explore the evolutions of the non-linearity parameter in the non slow-roll regime up to the end of inflation.

\subsection{Strategy for computing $f_{NL}, \tau_{NL}$ and $g_{NL}$}
Let us briefly discuss the necessary ingredients from \cite{Yokoyama:2007dw,Yokoyama:2008by} which are relevant for computing the non-linearity parameters. For a given generic form of the scalar potential, the concise expressions for the non-linearity parameters $f_{NL}, \tau_{NL}$ and $g_{NL}$ are \footnote{To avoid any possible confusion with the notations, we mention that all capital letters appearing as the sub/super scripts are denoted in calligraphic font.},
\begin{subequations}
\bea
\label{eq:fNL}
& & { f_{NL}} =\frac{5}{6}\, \frac{1}{ ({ N_{\cal D}^*} \,  { \Theta^{\cal D}_*})^2} \biggl[{ N_{{\cal A}{\cal B}}^F} \, { \Theta^{\cal A}_F} \,  { \Theta^{\cal B}_F} +\int_{N_*}^{N_F} { N_{\cal A}} \, { Q^{\cal A}_{{\cal B}{\cal C}}} \, { \Theta^{\cal B}} \, { \Theta^{\cal C}}  dN \biggr]
\eea
\bea
\label{eq:tauNL}
& & \hskip-3.5cm { \tau_{NL}} =\frac{1}{ ({ N_{\cal D}^*} \,  { \Theta^{\cal D}_*})^3} \biggl[A^{{\cal A}{\cal B}}\, \Omega_{\cal A}(N_*)\, \Omega_{\cal B}(N_*)\biggr]
\eea
\bea
\label{eq:gNL}
& & \hskip1cm  {g_{NL}} =\frac{25}{54}\, \frac{1}{ ({ N_{\cal D}^*} \,  { \Theta^{\cal D}_*})^3} \biggl[{ N_{{\cal A}{\cal B}{\cal C}}^F} \,{ \Theta^{\cal A}_F} \,  { \Theta^{\cal B}_F} \,{ \Theta^{\cal C}_F} +3\, \int_{N_*}^{N_F} { \Omega_{\cal A}} \, { Q^{\cal A}_{{\cal B}{\cal C}}} \, { \Theta^{\cal B}} \, { \Theta^{\cal C}}  dN \nonumber\\
& & \hskip4cm + \int_{N_*}^{N_F} { N_{\cal A}} \, {\cal Q}^{\cal A}_{{{\cal B}{\cal C}{\cal D}}} \, { \Theta^{\cal B}} \, { \Theta^{\cal C}} \, { \Theta^{\cal D}} dN \biggr]
\eea
\end{subequations}
Let us explain the meaning of the various symbols,
\begin{itemize}
\item{It is important to recall that in (\ref{eq:fNL},\ref{eq:tauNL},\ref{eq:gNL}), any object ${\cal O}^{\cal A}$ has two components ${\cal O}^{\cal A}\equiv \{{\cal O}^a_1, {\cal O}^a_2\}$ and the indices are appearing due to field redefinitions (\ref{eq:redef}), i.e.
$$ \varphi^a_1 \equiv \phi^a, \, \, \, \, \, \varphi^a_2 \equiv \frac{d \phi^a}{dt}, \,\,\,\,\ {\rm where}\,\, a \in \{1,\, 2\}. $$}
\item{The expressions of various derivatives $\{N_{\cal A}(N), \Theta^{{\cal A}}(N), \Omega_{{\cal A}}(N)\}$ at a generic time during the inflationary dynamics is required for computing the integrals. These vector quantities have to be computed by solving the respective ODEs. This is among the main task of the computation and we will elaborate more on it later.}
\item { The symbols ${ \Theta^{\cal A}}(N)$ are defined as \bea
& &  { \Theta^{\cal A}}(N) \equiv { \Lambda^{\cal A}_{\, \, {\cal B}}}(N, N_*) \, { N^{\cal B}_*} \,; \, \, \, \, \, {\rm where} \\
& &  { \Lambda^{\cal A}_{\, \, {\cal B}}}(N, N_*) = \left[T \exp\left(\int_{N^*}^{N} { P^{\cal A}_{\,\,\, {\cal B}}}(N) \, dN\right)\right] \, \, {\rm with} \, \,  { \Lambda^{\cal A}_{\, \, {\cal B}}}(N_*, N_*)= \delta^{\cal A}_{\cal B}. \nonumber
\eea
In the above, $T$ denotes time-ordering. The expressions for $P^{\cal A}_{\,\,\, {\cal B}}(N)$ are given as,
\bea
& & P^{a1}_{\, \, \, 1b}= -\frac{1}{6 H^3} \, \varphi^a_2 \, V_b\nonumber\\
& & P^{a1}_{\, \, \, 2b}= -\frac{V^a_{\, \, \, b}}{ H}+ \frac{1}{6 H^3} \, V^a V_b -R^a_{\, \,  \,c b d} \, \varphi^c_2 \, \varphi^d_2\\
& & P^{a2}_{\, \, \, 1b}= \frac{1}{H} \delta^a_b -\frac{1}{6 H^3} \, \varphi^a_2 \, ({\cal G}_{bd}\varphi^d_2) \nonumber\\
& & P^{a2}_{\, \, \, 2b}= -3 \, \delta^a_b + \frac{1}{6 H^3} \, V^a \, ({\cal G}_{bc}\varphi^c_2). \nonumber
\eea }
\item{The symbol $ {A^{{\cal A}{\cal B}}}$ is defined as,  \bea \left<\delta \varphi^{\cal A}_* \, \delta \varphi^{\cal B}_*\right> = { A^{{\cal A}{\cal B}}}\, \left(\frac{H_*}{2\pi}\right)^2 .\eea
In general, $ {A^{{\cal A}{\cal B}}}$ depends on the non-flat background metric. The respective expressions are given in appendix (\ref{expressions}) taking the slow-roll corrections \cite{Byrnes:2006vq} into account.}
\item{The expressions of various derivatives of e-folding $N$ evaluated at some final constant time-hypersurface $t_F$ (e.g. $N_{\cal A}^F, N_{{\cal A}{\cal B}}^F, N_{{\cal A}{\cal B}{\cal C}}^F$) are given as,
\bea
& & N_{\cal A}^F = -\left(\frac{H_{\cal A}}{H_{\cal D}\, F^{\cal D}}\right)_{at \, \, \varphi=\varphi^{(0)}(N_F)}\nonumber\\
& & N_{{\cal A}{\cal B}}^F = -\left(\frac{U_{{\cal A}{\cal B}}}{H_{\cal D}\, F^{\cal D}}\right)_{at \, \, \varphi=\varphi^{(0)}(N_F)}\\
& & N_{{\cal A}{\cal B}{\cal C}}^F = -\left(\frac{Z_{{\cal A}{\cal B}{\cal C}}}{H_{\cal D}\, F^{\cal D}}\right)_{at \, \, \varphi=\varphi^{(0)}(N_F)}\nonumber
\eea}
\item{The expressions for quantities $H_{\cal A}(N), H_{{\cal A}{\cal B}}(N), H_{{\cal A}{\cal B}{\cal C}}(N)$,$ U_{{\cal A}{\cal B}}(N)$,${ Z_{{\cal A} {\cal B}{\cal C}}}(N)$ as well as ${ Q^{\cal A}_{\,\,\,\, {\cal B}{\cal C}}}(N)$ and ${ {\cal Q}^{\cal A}_{\, \, \, {\cal B}{\cal C}{\cal D}}}(N)$ involve various derivatives of the scalar potential and the Hubble rate. Being quite lengthy, their explicit expressions can be found in (\ref{expressions}).}
 \end{itemize}
The main advantage of the formulation involving the redefinitions (\ref{eq:redef}) is that this simplifies the computation of the non-linearity parameters $f_{NL},\tau_{NL}$ and $g_{NL}$. It remains to solve first order ODEs for vector quantities (like $\Theta^{\cal A}(N)$, $ N_{\cal A}(N)$, $\Omega_{\cal A}(N)$). Furthermore, this formulation reduces the number of ${\cal O}(n^2)$ calculations to ${\cal O}(n)$ where $n$ is the number of scalar fields involved in the dynamics. This makes numerical calculations much more efficient in a multi-field scenario which has a large number of scalar fields. 

The expressions for $\Theta^{\cal A}(N)$, $ N_{\cal A}(N)$, $\Omega_{\cal A}(N)$ required for solving the integrals can be obtained by solving the following set of first order ODEs,
\bea
\label{eq:ODEs}
& & \frac{D}{dN} { N_{\cal A}}(N) = - { P^{\cal A}_{\, \, \, {\cal B}}}(N) \, { N_{\cal B}}(N), \nonumber\\
& & \frac{D}{dN} { \Theta^{\cal A}}(N) =  { P^{\cal A}_{\, \, \, {\cal B}}}(N) \, { \Theta^{\cal B}}(N),\\
& & \frac{D}{dN} { \Omega_{\cal A}}(N) = -\Omega_{\cal B}(N) \, { P^{\cal B}_{\, \, \, {\cal A}}}(N) -N_{\cal B}(N) \, Q^{\cal B}_{\,\,\,\, {\cal A}{\cal C}}(N)\, { \Theta^{\cal C}}(N). \nonumber
\eea 
Each set of these involves four ODEs for each quantity $\Theta^{\cal A}(N)$, $ N_{\cal A}(N)$ and $\Omega_{\cal A}(N)$. Of course, even this simplification of the problem into solving first order ODEs does not allow to proceed analytically for a given generic multi-field potential. However for concrete models, it is much easier to solve the aforementioned twelve ODEs numerically. As an important observation, the first two expressions of (\ref{eq:ODEs}) are mutually dual to each other. This implies that the quantity ${\cal X}(N_*) = N_{\cal A}(N_*)\, \Theta^{\cal A}(N_*)$ is constant irrespective of $N_*$. The algorithmic approach of solving the ODEs in (\ref{eq:ODEs}) can be summarized as,
\begin{itemize} 
 \item{First, one numerically solves the set of ODEs for ${ N_{\cal A}}(N)$ by using the boundary conditions corresponding to the final constant time-hypersurface $N_{{\cal A}}(N_F) = N_{{\cal A}}^F = - \left[\frac{H_{{\cal A}}}{ H_{\cal C} \, {\cal F}^{\cal C}}\right]_{N=N_F}$. Then, utilizing the numerical solutions obtained, one traces back to ${ N_{\cal A}}(N_*)$.}
\item{Using the backward traced values for ${ N_{\cal A}}(N_*)$, one gets the initial conditions $\Theta^{\cal A}(N_*) = A^{{\cal A}{\cal B}}(N_*) \, N_{\cal B}(N_*)$ and subsequently one numerically solves the set of ODEs for ${ \Theta^{\cal A}}(N)$.}
\item{Utilizing the solutions for $\Theta^{\cal A}(N)$, one traces forward for ${ \Theta^{\cal A}}(N_F)$, and then one can easily solve the set of ODEs for ${ \Omega_{\cal A}}(N)$ via using the initial conditions $\Omega_{\cal A}(N_F) = N_{{\cal A}{\cal B}}^F \, \Theta^{\cal B}(N_F)$.}
\end{itemize}
Substituting the various numerical solutions for all the relevant quantities in (\ref{eq:fNLgNLtauNLgen}), one obtains the values for all the three non-linearity parameter $f_{NL}, \, \tau_{NL}$ and $g_{NL}$.

\subsection{Numerical Results}
Now, we present the numerical results applying the strategy discussed so far for our potential (\ref{eq:Vinf}). The various non-linearity parameters $f_{NL}, \tau_{NL}$ and $g_{NL}$ have been estimated for slow-roll as well as beyond slow-roll regime for the four trajectories mentioned before corresponding to sufficiently large e-folding $N>50$. 
\begin{table}[ht]
  \centering
  \begin{tabular}{|c|c||c||c|c|c|}
  \hline
    $\tau_w$  & $\rho_w$  & $N$  & $|f_{NL}|$  & $g_{NL}$  & $\tau_{NL}$     \\
    \hline \hline
     3.9 & 1.495 & 96.51 & 0.020 & -0.010 & 6.26 $\times 10^{-4}$ \\
     \hline
    4 & 0.496 & 62.60 & 0.030 & -0.009 & 1.35 $\times 10^{-3}$ \\
   \hline
    4.5 & 1.475 & 51.17 & 0.035 &  -0.010 &  1.89 $\times 10^{-3}$ \\
    \hline
      5 & 1 & 60 & 0.017 & -0.0097 & 4.54 $\times 10^{-4}$ \\
  \hline
  \end{tabular}
  \caption{Various non-linearity parameters estimated up to the end of slow-roll region.}
  \label{tablesollow}
 \end{table}
We observe that in the slow-roll regime,  the $f_{NL}$ (as well as $\tau_{NL},  g_{NL}$) are very small which is also something quite expected \cite{Maldacena:2002vr}. Table \ref{tablesollow} presents the collection of non-linearity parameter values estimated in the slow-roll regime.
\begin{table}[ht]
  \centering
  \begin{tabular}{|c|c||c|c||c|c|c|c|}
  \hline
    $\tau_w$  & $\rho_w$   & $N^{\dagger}$& $|f_{NL}^{\dagger}|$  & $N_F$  & $|f_{NL}|$& $g_{NL}$  & $\tau_{NL}$     \\
    \hline \hline
     3.9 & 1.495 & 98.42 & 0.07 & 98.74 & 9.8 & 592.2 & 138.3 \\
     \hline
    4 & 0.496 & 64.67 & 0.08 & 65 & 377 & 4.39 $\times 10^8$ & 2.05 $\times 10^5$\\
    \hline
     4.5 & 1.475 & 53.08 & 0.09 & 53.42 & 48.6 & 4.27 $\times 10^5$ & 3404.5\\
    \hline
    5 & 1  & NULL & NULL &62 & 0.068 & -0.34 & 0.0067\\
    \hline   
  \end{tabular}
   \caption{Various non-linearity parameters estimated in the beyond slow-roll regime. The time corresponding to  $N^{\dagger}$ is simply  the time where axion $\rho_w$ first crosses its minimum and oscillates up to the end of inflation.}
   \label{tabelbeyond}
 \end{table}

Table \ref{tabelbeyond} shows that one can get large non-linearity parameters in the beyond slow-roll region for these trajectories except the single-field one. Especially, most of these large values are generated after the time when axion crosses its minimum for the first time, i,e. between $N_F$ and $N^{\dagger}$.
One can see that in the beyond slow-roll regime ($N_F-N=2$ or at most $3$), the trajectories acquire quick turns via oscillations in axionic directions when one approaches towards the end of inflation. 
This might be a reason for these non linearity parameters getting enhanced to significantly large values. Another reason for having large values of non-linearity parameters could be attributed to one of the slow-roll parameters getting significantly large, $\eta\sim{\cal O}(10^2-10^3)$. Moreover, contributions to the $\epsilon$ parameter coming from each of the two-fields are such that $\epsilon_1 \gg \epsilon_2$ in beyond slow-roll regime\footnote{In two-field model with $\epsilon_1 \gg \epsilon_2$, it has been argued that $f_{NL}$ can be enhanced by a factor of ${\cal O}\left(\epsilon_1/\epsilon_2\right)$ to its naively expected ${\cal O}(\epsilon)$ value \cite{Suyama:2010uj}. Note that such hierarchies in slow-roll parameters can be extremely crucial and this has been a key in realizing large $f_{NL}$ even in slow-roll regime \cite{Byrnes:2008wi}. Similar example could be a two-field DBI inflationary model like \cite{Cai:2009hw}.}. These can cause large enhancements in the intermediate quantities such as 
$\Theta^{\cal A}$ in the regime where the slow-roll condition is {\it strongly} violated. The reason could be simply thought of to be as $\Theta^{\cal A}(N) = \Lambda^{\cal A}_{\,\,\, \cal B}(N,N') N^{\cal B}(N')$ and $\Lambda^{\cal A}_{\,\,\, \cal B}(N,N')$ are defined through exponential of integrals 
involving $P^{\cal A}_{\cal B}(N)$ which depends on the combinations of slow-roll parameters. However, in the slow-roll regime, $\Lambda^{\cal A}_{\,\,\, \cal B}(N,N') \sim {\cal O}(1)$ and this results in the $f_{NL}$ parameters to be of slow-roll suppressed values for various trajectories.  Also, in \cite{Tanaka:2010km}, it has been argued that the value of $f_{NL}$ parameter can be expected to be crucially enhanced or suppressed if $\Lambda^{\cal A}_{\,\,\, \cal B}(N,N')$ happens to be so.

\section{Conclusions and Discussions}
\label{sec_Conclusions and Discussions}

In this article, we generalized the standard single field poly-instanton inflationary model to a two-field model in which inflation is driven by a combined dynamics of a (Wilson) divisor volume mode and the respective $C_4$ axion. In this setup, we have focused on two aspects. The first one was the study related to the two-field inflationary model. The second one has been an investigation of the possibility to realize the non-Gaussianities by computing the non-linearity parameters such as $f_{NL}, \tau_{NL}$ and $g_{NL}$. This was done in the slow-roll as well as in the beyond slow-roll regime. 

In the context of the first aspect, we studied the evolutions of background fields and explored the various possible types of inflationary trajectories. We observed that, depending on the choice of initial conditions, one can have inflationary trajectories corresponding to a wide range (order one e-folding to quite large) number of e-foldings. However, we mainly focused on the class of trajectories which could produce order 50 (or more) e-foldings. The nature of various trajectories are also quite different; some are attracted (repelled) inwards (outwards) to the attractor point in a straight line and hence implying that such trajectories have no isocurvature perturbations. The other type of trajectories have significant curving due to the axion dynamics and do have isocurvature perturbations. Further, if the axion initial conditions are not very far from the respective minimum, the respective trajectories are attracted into the nearest valley. Interestingly, there are some trajectories which are 
predominantly axionic before getting trapped into an attractor point. This situation is similar to the rotation of a roulette ball before being trapped into a particular slot, and so justifies the name ``roulette'' inflation \cite{Bond:2006nc}. We analyzed the running of the slow-roll parameters, the Hubble rate and the power spectrum of scalar perturbations in terms of e-foldings. Similar to the single-field inflationary models \cite{Conlon:2005jm,Blumenhagen:2012ue} realized in type IIB LARGE volume orientifold compactifications, the values of slow-roll parameters $\epsilon$ and $\eta$ are quite hierarchial. The slow-roll parameter $\epsilon$ is very small ($\epsilon \sim 10^{-9}$) during slow-roll and it increases only up to order $10^{-6}$ values when slow-roll conditions are violated by $\eta \sim {\cal O}(1)$. In fact, the $\epsilon$ parameter increases exponentially fast near the end of inflation and at the same time the $\eta$ parameter also 
gets enhanced to a very large values. Thus, from the point of view of e-foldings, a very narrow window is available in the `beyond' slow-roll regime in which $\eta$ parameter is very large. Moreover, a significant hierarchy $\epsilon_1 \gg \epsilon_2$ is also realized within the components of $\epsilon$ in beyond slow-roll regime. Also, as we have found several trajectories with significant curving, it has been a good motivation for looking at the non-Gaussianities signatures in this ``roulette poly-instanton inflation'' setup.  

Subsequently, we have investigated the possibilities for generating detectable values for the non-linearity parameters $f_{NL}, \tau_{NL}$ and $g_{NL}$ for various trajectories. We followed the strategy developed in \cite{Yokoyama:2007dw,Yokoyama:2008by} which is applicable for a given {\it generic} scalar potential and is also valid in the beyond slow-roll regime. For the trajectory which has initial conditions such that axion sits at its minimum position to start with, this two-field inflationary process reduces to its single-field analogue \cite{Blumenhagen:2012ue}. Thus, for this trajectory, we get quite small value for the non-linearity parameters $f_{NL}$, $\tau_{NL}$ and $ g_{NL}$ in the slow-roll regime. In fact, within the slow-roll regime, we obtain small values for each of the the three non-linearity parameters for all the `types' of trajectories we have classified in this setup. However, in the beyond slow-roll regime ($N_F-N=2$ or at most $3$), the trajectories acquire turns due to oscillations 
in axionic 
directions when one approaches towards the end of inflation.
These might be responsible for the non linearity parameters getting enhanced to significantly large values. Moreover, a sharp increment of the $\eta$ parameter towards the end of inflation could be another reason for the non-linearity parameters getting enhanced in the beyond slow-roll regime. Recently, in the context of estimating non-Gaussianities signatures in the beyond slow-roll regime, large $f_{NL}$ values have been observed in \cite{Byrnes:2009qy,Battefeld:2009ym} with a potential having a combination of various exponential terms such that the Hubble rate is sum separable. Such a potential has been argued to be possibly realized in string models such as \cite{Conlon:2005jm}. Although, our two-field scalar potential does not have a separable form for the Hubble rate (and hence one can not 
expect to reproduce our results  from the strategy developed in  \cite{Byrnes:2009qy,Battefeld:2009ym}), nevertheless there are important similarities which could be responsible for the large values of $f_{NL}$ parameter. A few of these are,
\begin{itemize}
\item{Large $f_{NL}$ is  realized {\it only} in the region of the final stage of inflation where there is no significant increment in the number of e-folds. }
\item{In both of the cases, there is a significant enhancement in the $\eta$-parameter towards the end of the inflation resulting in hierarchial values for the $\epsilon$ and $\eta$ parameters.}
\item{Without going into all the details of  \cite{Battefeld:2009ym}, we state as an observation that the exponent appearing in the form of the Hubble rate as $e^{-\alpha_k \, \phi_k}$  results in $f_{NL} \sim {\cal O}(n \, {\alpha_k} /\sqrt{m_k}) $, where $n$ is the number of fields involved in the dynamics and $\alpha_k, \, m_k$ are some model dependent parameters. For $\alpha_k \sim {\cal O}(100), m_k \sim 1$, one gets $f_{NL} \sim {\cal O}(100)$.  Now for our case, after considering the canonically normalized forms of the divisor volume modulus, as seen in single-field analogue in \cite{Blumenhagen:2012ue}, the exponential term in the potential appears as $e^{-\alpha_k \, \phi_k^{4/3}}$ where $\alpha_k \sim{ \ov{\cal V}^{2/3}} $ and $\ov{\cal V}$ is the stabilized value for the Calabi Yau volume. For the present case, $\ov{\cal V} \sim 905$.} Moreover, recall that such exponential form results in $\eta \sim \sqrt\epsilon \sim  e^{- \alpha_k \, \phi_k^{4/3}}$
 which is similar to those of \cite{Battefeld:2009ym}.
\end{itemize}   
Along these lines, our analysis could be a supporting step towards looking for the non-Gaussianities in the beyond slow-roll limit. Also, a systematic study of what happens after the end of inflation, for example following the motion/decay of inflaton fields and  reheating issues etc. can be extremely crucial in such models\footnote{We thank E. Copeland and A. Mazumdar for pointing this out to us and bringing our attention to the related studies in \cite{Mazumdar:2010sa,Cicoli:2010ha,Cicoli:2010yj,Leung:2012ve}}. In the context of K\"ahler moduli inflation models \cite{Conlon:2005jm,Cicoli:2008gp}, it has been found that there is a possibility of inflaton dumping all its energy into the hidden sector instead of the visible sector \cite{Cicoli:2010ha,Cicoli:2010yj}. For investigating this aspect, one has to extend our setup by {explicitly} embedding the visible sector. It would be interesting to know how generic is the problem and we hope to get back with this aspect in a future work.

\subsubsection*{Acknowledgments}
We would like to thank Ralph Blumenhagen, Godfrey Leung, Anupam Majumdar, Aalok Misra, Takahiro Tanaka, Lingfei Wang and Shuichiro Yokoyama for helpful discussions. In addition, we also gratefully acknowledge very enlightening discussions and comments from Shuichiro Yokoyama and Lingfei Wang. PS would like to thank the IPhT Saclay group and the Institut Henri Poincar\'e (IHP), Paris for hospitality during  visits to the respective centers (during SCGSC-2012) where related inflationary aspects have been presented.  XG is supported by the MPG-CAS Joint Doctoral Promotion Programme and PS is supported by a postdoctoral research fellowship from the Alexander von Humboldt Foundation.
\begin{appendix}
\section{Collection of the relevant expressions}
\label{expressions}
Recalling the field redefinitions,
$$\varphi^a_1 = \phi^a, \, \varphi^a_2 = \frac{d\phi^a}{dt}$$
the background field-evolution is governed by the following equivalent expressions,
$$\frac{d}{dN} \varphi^a_1 \equiv F^a_1=\frac{1}{H}\,\varphi^a_2, \, \, \, \frac{D}{dN}\varphi^a_2 \equiv F^a_2 = -3 \varphi^a_2 - \frac{{\cal G}^{ab}\, V_b}{H}$$
where $D$ is the covariant derivative defined as $D \varphi^a_2 = d \varphi^a_2 + {\Gamma^a}_{bc} \varphi^b_2  d\varphi^c_1$, and Hubble rate is defined via 
$H^2=\f{1}{3}\left(V+\f{1}{2}\, {\cal G}_{ab} \varphi^a_2 \varphi^b_2 \right)$. Now, utilizing the aforementioned relations, the various useful expressions appearing at the intermediate stages in the computation of non-linearity parameters ($f_{NL}, \tau_{NL}$ and $g_{NL}$) can be easily derived. The expressions for $P^{\cal A}_{\,\,{\cal B}} = \left(\frac{D F^{\cal A}}{\partial \varphi^{\cal B}}\right)_{at \, \, \varphi^{\cal A}=\varphi^{\cal A}_{(0)}(N)}$ are simply given as
\bea
& & P^{a1}_{\, \, \, 1b}= -\frac{1}{6 H^3} \, \varphi^a_2 \, V_b\nonumber\\
& & P^{a1}_{\, \, \, 2b}= -\frac{V^a_{\, \, \, b}}{ H}+ \frac{1}{6 H^3} \, V^a V_b -R^a_{\, \,  \,c b d} \, \varphi^c_2 \, \varphi^d_2\\
& & P^{a2}_{\, \, \, 1b}= \frac{1}{H} \delta^a_b -\frac{1}{6 H^3} \, \varphi^a_2 \, ({\cal G}_{bd}\varphi^d_2) \nonumber\\
& & P^{a2}_{\, \, \, 2b}= -3 \, \delta^a_b + \frac{1}{6 H^3} \, V^a \, ({\cal G}_{bc}\varphi^c_2) \nonumber
\eea
The expressions for $Q^{\cal A}_{\,\,{\cal B}{\cal C}} = \left(\frac{D P^{\cal A}_{\cal B}}{\partial \varphi^{\cal C}}\right)_{at \, \, \varphi^{\cal A}=\varphi^{\cal A}_{(0)}(N)}$ are given as
\bea
& & Q^{a11}_{\, \, \,  1bc}= -\frac{1}{H^3}\, \varphi^a_2 \, V_{bc} + \frac{1}{12 \, H^5} \, \varphi^a_2 \, V_b \, V_c - R^a_{\, \,  \,bcd} \,\varphi^d_2\nonumber\\
& & Q^{a12}_{\, \, \,  1bc}= -\frac{1}{H^3}\, \delta^a_{\, \, \, c} \, V_b + \frac{1}{12 \, H^5} \, \varphi^a_2 \, V_b ({\cal G}_{cd}\varphi^d_2) \nonumber\\
& & Q^{a21}_{\, \, \,  1bc}= -\frac{1}{H^3}\, \delta^a_{\, \, \, b} \, V_c + \frac{1}{12 \, H^5} \, \varphi^a_2 \, ({\cal G}_{bd}\varphi^d_2) V_c \nonumber\\
& & Q^{a22}_{\, \, \,  1bc}= -\frac{1}{H^3}\, \delta^a_{\, \, \, b} \, ({\cal G}_{cd}\varphi^d_2) + \frac{1}{12 \, H^5} \,\varphi^a_2 \, ({\cal G}_{bd}\varphi^d_2)({\cal G}_{cf}\varphi^f_2) \nonumber\\
& & \hskip 3cm -\frac{1}{H^3}\, \delta^a_{\, \, \, c} \, ({\cal G}_{bd}\varphi^d_2)-\frac{1}{H^3}\,\varphi^a_2 \, {\cal G}_{bc}\\
& & Q^{a11}_{\, \, \,  2bc}= -\frac{1}{H}\, V^a_{\, \, \, \, b c} + \frac{1}{6 H^3} \left(V^a_{\, \, \, b} \, V_c + V^a_{\, \, \, c} \, V_b +V^a \, V_{bc}\right) \nonumber\\
& & \hskip3cm -\frac{1}{12\, H^5} \, V^a V_b V_c - \nabla_c R^a_{\, \,  \,p b q} \,\varphi^p_2 \varphi^q_2\nonumber\\
& & Q^{a12}_{\, \, \,  2bc}= \frac{1}{6 \, H^3} \,V^a_{\, \, \, b} \,({\cal G}_{cd}\varphi^d_2)  -\frac{1}{12\, H^5} \, V^a \, V_b \, ({\cal G}_{cd}\varphi^d_2) -2 R^a_{\, \,  \,cbd} \, \varphi^d_2 \nonumber\\
& & Q^{a21}_{\, \, \,  2bc}= \frac{1}{6 \, H^3} \,V^a_{\, \, \, c} \,({\cal G}_{bd}\varphi^d_2)  -\frac{1}{12\, H^5} \, V^a \, ({\cal G}_{bd}\varphi^d_2)\, V_c -  R^a_{\, \,  \,dcb} \, \varphi^d_2 \nonumber\\
& & Q^{a22}_{\, \, \,  2bc}= -\frac{1}{12\, H^5} \, V^a \, ({\cal G}_{bd}\varphi^d_2)\,({\cal G}_{bd}\varphi^d_2) + \frac{1}{6 \, H^3} \, V^a \, {\cal G}_{bc}\nonumber
\eea
After including the slow-roll corrections in $A^{{\cal A}{\cal B}}$ \cite{Byrnes:2006vq} and keeping in mind that background metric for our case is diagonal, we get the following components,
\bea
\label{Aab}
& &  { A^{aa}_{11}} = {\cal G}^{aa} \biggl[1+2 \epsilon + \alpha\left( -4 \epsilon + {\cal G}_{aa}\biggl(\frac{d\phi^a}{dN}\right)^2-{\cal G}_{bb}\left(\frac{d\phi^b}{dN}\right)^2 - 2 \frac{{\cal G}^{aa} V_{aa}}{V}\biggr)\biggr] ;\nonumber\\
& & { A^{ab}_{11}} = 0\, \,\,\, {\rm for} \, \, \, a \ne b , \, \, {\rm where} \, \, a = \{1 , \, 2\} \, \, {\rm and} \, \, \alpha \sim 0.7 , \nonumber
\eea
\bea
& &  { A^{ab}_{12}} =\frac{{\cal G}^{ac} V_c \, {\cal G}^{bd} V_d}{V^2}-\frac{{\cal G}^{ac}{\cal G}^{bd} \left({V}_{cd}-\Gamma^q_{\,\, cd} V_q \right)}{V} = { A^{ab}_{21}} ,\\
& &  { A^{ab}_{22}} = \left(\frac{ V^a \, V_c}{V^2}-\frac{{\cal G}^{ak} \left({V}_{ck}-\Gamma^q_{\,\, ck} V_q \right)}{V} \right) \left(\frac{ V^c \, V^b}{V^2}-\frac{{\cal G}^{cp}{\cal G}^{bq} \left({V}_{pq}-\Gamma^l_{\,\, pq} V_l \right)}{V} \right). \nonumber
\eea
In the first two-lines of (\ref{Aab}), the indices are not summed over. Furthermore, the expressions for $H_{\cal A}, \, H_{{\cal A}{\cal B}},  H_{{\cal A}{\cal B}{\cal C}} $ are elaborated as,
\bea
& & H_a^1 = \frac{1}{6 \, H} \, V_a, \, \, H_a^2 = \frac{1}{6 \, H} \, ({\cal G}_{ab}\varphi^b_2) \\
& & H_{ab}^{11} = \frac{1}{6 \, H }\, V_{ab} - \frac{1}{36 \, H^3} \, V_a \, V_b \nonumber\\
& & H_{ab}^{12} =  - \frac{1}{36 \, H^3} \, V_a \, ({\cal G}_{bc}\varphi^c_2) \nonumber\\
& & H_{ab}^{21} =  - \frac{1}{36 \, H^3} \, V_b \, ({\cal G}_{ac}\varphi^c_2)\nonumber\\
& & H_{ab}^{22} =  \frac{1}{6 \, H}\, {\cal G}_{ab} - \frac{1}{36 \, H^3} \, ({\cal G}_{ac}\varphi^c_2) ({\cal G}_{bd}\varphi^d_2)\,\nonumber\\
& & H_{abc}^{111} = \frac{1}{6 \, H}\, V_{abc} -\frac{1}{36 \, H^3}\,\left(V_{ab} V_c +V_{ac} V_b+ V_a V_{bc}\right) + \frac{1}{72 \, H^5}\, V_a V_b V_c\nonumber\\
& & H_{abc}^{112} = -\frac{1}{36\, H^3} \, V_{ab} ({\cal G}_{cd}\varphi^d_2) + \frac{1}{72 \, H^5}\,V_a V_b ({\cal G}_{cd}\varphi^d_2)\nonumber\\
& & H_{abc}^{121} = \frac{1}{72 \, H^5}\,V_a  ({\cal G}_{bd}\varphi^d_2) V_c-\frac{1}{36\, H^3} \, V_{ac} ({\cal G}_{bd}\varphi^d_2)\nonumber\\
& & H_{abc}^{122} = \frac{1}{72 \, H^5}\,V_a  ({\cal G}_{bd}\varphi^d_2) ({\cal G}_{cp}\varphi^p_2)-\frac{1}{36\, H^3} \, V_{a} \, {\cal G}_{bc}\nonumber\\
& & H_{abc}^{211} = \frac{1}{72 \, H^5}\, ({\cal G}_{ad}\varphi^d_2) \, V_b V_c -\frac{1}{36\, H^3} \, ({\cal G}_{ad}\varphi^d_2) \, V_{bc} \nonumber\\
& & H_{abc}^{212} = \frac{1}{72 \, H^5}\,({\cal G}_{ad}\varphi^d_2) V_b \, ({\cal G}_{cp}\varphi^p_2)-\frac{1}{36\, H^3} \, {\cal G}_{ac}\, V_{b} \nonumber\\
& & H_{abc}^{221} = -\frac{1}{36\, H^3} \, {\cal G}_{ab}\, V_c + \frac{1}{72 \, H^5}\,({\cal G}_{ad}\varphi^d_2) \, ({\cal G}_{bp}\varphi^p_2) \, V_c \nonumber\\
& & H_{abc}^{222} = \frac{1}{72 \, H^5}\,({\cal G}_{ad}\varphi^d_2) ({\cal G}_{bp}\varphi^p_2) \, ({\cal G}_{cq}\varphi^q_2)-\frac{1}{36\, H^3} \, {\cal G}_{ab}\, ({\cal G}_{cd}\varphi^d_2)\nonumber\\
& & \hskip 3.99cm -\frac{1}{36\, H^3} \, {\cal G}_{ac}\, ({\cal G}_{bd}\varphi^d_2) -\frac{1}{36\, H^3} \, ({\cal G}_{ad}\varphi^d_2)\, {\cal G}_{bc}\nonumber
\eea
The expressions of various derivatives of e-folding $N$ evaluated at a final time-hypersurface $t_F$ (e.g. $N_{\cal A}^F, N_{{\cal A}{\cal B}}^F, N_{{\cal A}{\cal B}{\cal C}}^F$) which is used for providing the initial conditions while solving for the ODEs backward in time are given as,
\bea
& & N_{\cal A}^F = -\left(\frac{H_{\cal A}}{H_{\cal D}\, F^{\cal D}}\right)_{at \, \, \varphi=\varphi^{(0)}(N_F)}\nonumber
\eea
\bea
& & N_{{\cal A}{\cal B}}^F = -\left(\frac{U_{{\cal A}{\cal B}}}{H_{\cal D}\, F^{\cal D}}\right)_{at \, \, \varphi=\varphi^{(0)}(N_F)}\\
& & N_{{\cal A}{\cal B}{\cal C}}^F = -\left(\frac{Z_{{\cal A}{\cal B}{\cal C}}}{H_{\cal D}\, F^{\cal D}}\right)_{at \, \, \varphi=\varphi^{(0)}(N_F)}\nonumber
\eea
where
\bea
& & U_{{\cal A}{\cal B}} = H_{{\cal A}{\cal B}} + 2 \left(H_{\cal C} \, P^{\cal C}_{\cal A} + F^{\cal C} \, H_{{\cal C}{\cal A}} \right) N_{\cal B}^F  \nonumber\\
& & \hskip3cm + \left(F^{\cal C} H_{{\cal C}{\cal D}}\, F^{\cal D} + H_{\cal C} \, P^{\cal C}_{\cal D} \, F^{\cal D}\right)N_{\cal A}^F \, N_{\cal B}^F \nonumber\\
& & \hskip-0.2cm Z_{{\cal A}{\cal B}{\cal C}} = H_{{\cal A}{\cal B}{\cal C}}+\biggl[H_{\cal D}\left(Q^{\cal D}_{{\cal E}{\cal F}} \, F^{\cal E} + P^{\cal D}_{\cal E}\, P^{\cal E}_{\cal F}\right)F^{\cal F} + H_{{\cal D}{\cal E}{\cal F}}\, F^{\cal D} F^{\cal E} \, F^{\cal F}\\
& & \hskip0.8cm + 3 \, F^{\cal D} H_{{\cal D}{\cal E}}\, P^{\cal E}_{\cal F} \, F^{\cal F}\biggr] N_{\cal A}^F N_{\cal B}^F N_{\cal C}^F + 3 \biggl[\left(H_{{\cal A}{\cal D}{\cal E}}\, F^{\cal D} + H_{{\cal A}{\cal D}}\, P^{\cal D}_{\cal E}\right)F^{\cal E} \nonumber\\
& & \hskip0.8cm + 2 \, F^{\cal D} H_{{\cal D} {\cal E}}\, P^{\cal E}_{\cal A} + H_{{\cal D}}\left(Q^{\cal D}_{{\cal E}{\cal A}}\, F^{\cal E} + P^{\cal D}_{\cal E}\, P^{\cal E}_{\cal A}\right)\biggr] N_{\cal B}^F \, N_{\cal C}^F \nonumber\\
& & \hskip0.8cm + 3\, \left(2 H_{{\cal A}{\cal D}}\,P^{\cal D}_{\cal B} + F^{\cal D}\,H_{{\cal D}{\cal A}{\cal B}}\,+H_{\cal D}\, Q^{\cal D}_{{\cal A}{\cal B}}\right)\, N_{\cal C}^F\nonumber\\
& & \hskip0.8cm + 3\, \left(F^{\cal D}\, H_{{\cal D}{\cal E}}\, F^{\cal E} + H_{\cal D} \, P^{\cal D}_{\cal E}\, F^{\cal E}\right) N_{\cal A}^F\, N_{{\cal B}{\cal C}}^F + 3 \left(F^{\cal D}\, H_{{\cal D}{\cal A}}+ H_{\cal D}\, P^{\cal D}_{\cal A}\right)\, N_{{\cal B}{\cal C}}^F\nonumber
\eea
Finally, the expressions for ${\cal Q}^{\cal A}_{\,\,{\cal B}{\cal C}{\cal D}} = \left(\frac{D Q^{\cal A}_{{\cal B}{\cal C}}}{\partial \varphi^{\cal D}}\right)_{at \, \, \varphi^{\cal A}=\varphi^{\cal A}_{(0)}(N)}$ are given as,
\bea
& & {\cal Q}^{a111}_{\, \, \,  1bcd} = \frac{1}{12 H^5} \, \varphi^a_2  \, V_{bc} \, ({\cal G}_{dp}\varphi^p_2) - \frac{1}{6 H^3} \,\delta^a_{\, \, \, d} \, V_{bc} \nonumber\\
& & \hskip 3cm  -\frac{5}{72 H^7} \,\varphi^a_2 \, V_b \, V_c \, ({\cal G}_{dp}\varphi^p_2) + \frac{1}{12 H^5} \, \delta^a_{\, \, \, d} \, V_b \, V_c \nonumber\\
& & {\cal Q}^{a112}_{\, \, \,  1bcd} = -\frac{1}{6 H^3} \, \delta^a_{\,\,\, d} \, V_{bc} + \frac{1}{12 H^5} \, \delta^a_{\, \, \, d} \, V_b \, V_c \nonumber\\
& & \hskip 3cm -\frac{5}{72 H^7}\, \varphi^a_2 \, V_b \,V_c\, ({\cal G}_{dp}\varphi^p_2) +\frac{1}{12 H^5} \, \varphi^a_2  \, V_{bc} \, ({\cal G}_{dp}\varphi^p_2)\nonumber\\
& & {\cal Q}^{a121}_{\, \, \,  1bcd} = -\frac{1}{6 H^3} \, \delta^a_{\,\,\, c} \, V_{bd} + \frac{1}{12 H^5} \, \delta^a_{\, \, \, c} \, V_b \, V_d \nonumber\\
& & \hskip 3cm -\frac{5}{72 H^7}\, \varphi^a_2 \, V_b \, ({\cal G}_{cp}\varphi^p_2) \, V_d +\frac{1}{12 H^5} \, \varphi^a_2  \, V_{bd} \, ({\cal G}_{cp}\varphi^p_2)\nonumber\\
& & {\cal Q}^{a211}_{\, \, \,  1bcd} = \frac{1}{12 H^5} \, \delta^a_{\, \, \, b} \, V_c \, V_d-\frac{1}{6 H^3} \, \delta^a_{\,\,\, b} \, V_{cd}  \nonumber\\
& & \hskip 3cm  +\frac{1}{12 H^5} \, \varphi^a_2  \,({\cal G}_{bp}\varphi^p_2)\, V_{cd}  -\frac{5}{72 H^7}\, \varphi^a_2 \,({\cal G}_{bp}\varphi^p_2)\, V_c \, V_d\nonumber\\
& & {\cal Q}^{a122}_{\, \, \,  1bcd} = \frac{1}{12 H^5} \, \delta^a_{\, \, \, b} \, V_c \,({\cal G}_{dp}\varphi^p_2)+\frac{1}{12 H^5} \, \delta^a_{\, \, \, d} \, V_b\, ({\cal G}_{cp}\varphi^p_2)  \nonumber\\
& & \hskip 3cm    -\frac{5}{72 H^7}\, \varphi^a_2 \,V_b \, ({\cal G}_{cp}\varphi^p_2)\,({\cal G}_{dq}\varphi^q_2) +\frac{1}{12 H^5} \, \varphi^a_2 \, V_b \, {\cal G}_{cd}\nonumber
\eea
\bea
& & {\cal Q}^{a221}_{\, \, \,  1bcd} = \frac{1}{12 H^5} \, \varphi^a_2 \, {\cal G}_{bc}\,  V_d+\frac{1}{12 H^5} \, \delta^a_{\, \, \, b} \, \,({\cal G}_{cp}\varphi^p_2) \, V_d   \nonumber\\
& & \hskip 3cm   +\frac{1}{12 H^5} \, \delta^a_{\, \, \, c} \, ({\cal G}_{bp}\varphi^p_2) \, V_d -\frac{5}{72 H^7}\, \varphi^a_2 \, ({\cal G}_{bp}\varphi^p_2) \, ({\cal G}_{cq}\varphi^q_2)\,V_d  \nonumber\\
& & {\cal Q}^{a212}_{\, \, \,  1bcd} = \frac{1}{12 H^5} \, \delta^a_{\, \, \, b} \, V_c \,({\cal G}_{dp}\varphi^p_2)+\frac{1}{12 H^5} \, \delta^a_{\, \, \, d} \, ({\cal G}_{bp}\varphi^p_2) \, V_c \nonumber\\
& & \hskip 3cm    -\frac{5}{72 H^7}\, \varphi^a_2 \,({\cal G}_{bp}\varphi^p_2) \, V_c\,({\cal G}_{dq}\varphi^q_2) +\frac{1}{12 H^5} \, \varphi^a_2 \, V_b \, {\cal G}_{cd}\nonumber\\
& & {\cal Q}^{a222}_{\, \, \,  1bcd} = \frac{1}{12 H^5}\biggl[ \delta^a_{\, \, \, b} \, ({\cal G}_{cp}\varphi^p_2)\, ({\cal G}_{dq}\varphi^q_2)+\delta^a_{\, \, \, c} \, ({\cal G}_{bp}\varphi^p_2)\, ({\cal G}_{dq}\varphi^q_2) +\varphi^a_2 \, {\cal G}_{bc} \, ({\cal G}_{dq}\varphi^q_2) \biggr]\nonumber\\
& & \hskip1.2cm -\frac{1}{6 H^3}\biggl[\delta^a_{\,\,\, b} \, {\cal G}_{cd}+\delta^a_{\,\,\, c} \, {\cal G}_{bd}+\delta^a_{\,\,\, d} \, {\cal G}_{bc}\biggr]- \frac{5}{72 H^7}\,\varphi^a_2\,({\cal G}_{bp}\varphi^p_2) ({\cal G}_{cr}\varphi^r_2)({\cal G}_{ds}\varphi^s_2) \nonumber\\
& & \hskip2.5cm +\frac{1}{12 H^5}\biggl[\delta^a_{\,\,\,d} ({\cal G}_{bq}\varphi^q_2)({\cal G}_{cr}\varphi^r_2) + \varphi^a_2 \, {\cal G}_{bd} ({\cal G}_{cr}\varphi^r_2) + \varphi^a_2 \, ({\cal G}_{br}\varphi^r_2)\, {\cal G}_{cd}\biggr]\nonumber\\
& & {\cal Q}^{a111}_{\, \, \,  2bcd} = -\frac{1}{H}\, V^a_{\, \, \, \, bcd} + \frac{1}{6 H^3} \biggl[V^a_{\, \, \, \, bc} \, V_d + V^a_{\, \, \, \, bd} \, V_c + V^a_{\, \, \, \, b} \, V_{cd}+V^a_{\, \, \, \, cd} \, V_b+V^a_{\, \, \, \,c} \, V_{bd}\nonumber\\
& & \hskip1.7cm +V^a_{\, \, \, \, d} \, V_{bc}+V^a \, V_{bcd}\biggr] -\frac{1}{12 H^5}\biggl[V^a_{\, \, \, \,b} \, V_{c} \, V_d +V^a_{\, \, \, \,c} \, V_{b} \, V_d +V^a_{\, \, \, \,d} \, V_{b} \, V_c \nonumber\\
& & \hskip1.7cm +V^a \, V_{bc} \, V_d +V^a \, V_{bd} \, V_c +V^a \, V_{b} \, V_{cd}\biggr]+ \frac{5}{72 H^7} \, V^a \, V_b \, V_c \, V_d\nonumber\\
& & {\cal Q}^{a112}_{\, \, \,  2bcd} = \frac{1}{6 H^3}\, V^a_{\, \, \, \, bc}\, ({\cal G}_{dq}\varphi^q_2)-\frac{1}{12 H^5}\, V^a_{\, \, \, \, b}\, V_c \, ({\cal G}_{dq}\varphi^q_2)-\frac{1}{12 H^5}\, V^a_{\, \, \, \, c}\, V_b \, ({\cal G}_{dq}\varphi^q_2)\nonumber\\
& & \hskip3cm -\frac{1}{12 H^5}\, V^a\, V_{bc} \, ({\cal G}_{dq}\varphi^q_2) +\frac{5}{72 H^7}\, V^a\,V_b\, V_c \, ({\cal G}_{dq}\varphi^q_2) \nonumber\\
& & {\cal Q}^{a121}_{\, \, \,  2bcd} = \frac{1}{6 H^3}\, V^a_{\, \, \, \, bd}\, ({\cal G}_{cq}\varphi^q_2)-\frac{1}{12 H^5}\, V^a_{\, \, \, \, b} \, ({\cal G}_{cq}\varphi^q_2)\, V_d +\frac{5}{72 H^7}\, V^a\,V_b\, ({\cal G}_{cq}\varphi^q_2) \, V_d \nonumber\\
& & \hskip3cm -\frac{1}{12 H^5}\, V^a_{\, \, \, \, d}\, V_b \, ({\cal G}_{cq}\varphi^q_2) -\frac{1}{12 H^5}\, V^a\, V_{bd} \, ({\cal G}_{cq}\varphi^q_2)  \nonumber\\
& & {\cal Q}^{a211}_{\, \, \,  2bcd} = \frac{1}{6 H^3}\, V^a_{\, \, \, \, cd}\, ({\cal G}_{bq}\varphi^q_2)-\frac{1}{12 H^5}\, V^a_{\, \, \, \, c} \, ({\cal G}_{bq}\varphi^q_2)\, V_d -\frac{1}{12 H^5}\, V^a \, ({\cal G}_{bq}\varphi^q_2)\, V_{cd}  \nonumber\\
& & \hskip3cm +\frac{5}{72 H^7}\, V^a\, ({\cal G}_{bq}\varphi^q_2)\,V_c \, V_d -\frac{1}{12 H^5}\, V^a_{\, \, \, \, d} \, ({\cal G}_{bq}\varphi^q_2) \, V_c  \nonumber\\
& & {\cal Q}^{a122}_{\, \, \,  2bcd} = \frac{1}{6 H^3}\, V^a_{\, \, \,\, b}\, {\cal G}_{cd} - \frac{1}{12 H^5}\, V^a_{\,\,\,\,b} \, ({\cal G}_{cp}\varphi^p_2)\, ({\cal G}_{dq}\varphi^q_2) \nonumber\\
& & \hskip3cm -\frac{1}{12 H^5}\, V^a \, V_b \, {\cal G}_{cd} +\frac{5}{72 H^7}\, V^a\, V_{b} \, ({\cal G}_{cp}\varphi^p_2)\, ({\cal G}_{dq}\varphi^q_2)\nonumber\\
& & {\cal Q}^{a221}_{\, \, \,  2bcd} =  -\frac{1}{12 H^5}\, V^a \,{\cal G}_{bd}\, V_c  - \frac{1}{12 H^5}\, V^a_{\,\,\,\,c} \, ({\cal G}_{bp}\varphi^p_2)\, ({\cal G}_{dq}\varphi^q_2) \nonumber\\
& & \hskip3cm   +\frac{5}{72 H^7}\, V^a \, ({\cal G}_{bp}\varphi^p_2)\, V_{c}\, ({\cal G}_{dq}\varphi^q_2)+ \frac{1}{6 H^3}\, V^a_{\, \, \,\, c}\, {\cal G}_{bd}\nonumber\\
& & {\cal Q}^{a212}_{\, \, \,  2bcd} =  - \frac{1}{12 H^5}\, V^a_{\,\,\,\,c} \, ({\cal G}_{bp}\varphi^p_2)\, ({\cal G}_{dq}\varphi^q_2) +\frac{5}{72 H^7}\, V^a\, ({\cal G}_{bp}\varphi^p_2)\, V_{c} \, ({\cal G}_{dq}\varphi^q_2) \nonumber\\
& & \hskip3cm +\frac{1}{6 H^3}\, V^a_{\, \, \,\, c}\, {\cal G}_{bd} -\frac{1}{12 H^5}\, V^a \, {\cal G}_{bd}\, V_c  \nonumber\\
& & {\cal Q}^{a222}_{\, \, \,  2bcd} = \frac{5}{72 H^7}\, V^a\, ({\cal G}_{bp}\varphi^p_2)\,({\cal G}_{cq}\varphi^q_2)\, ({\cal G}_{dr}\varphi^r_2) -\frac{1}{12 H^5} \, V^a \, {\cal G}_{bd}\, ({\cal G}_{cp}\varphi^p_2) \\
& & \hskip4cm -\frac{1}{12 H^5} \, V^a \, ({\cal G}_{bp}\varphi^p_2)\, {\cal G}_{cd}\, -\frac{1}{12 H^5} \, V^a \, {\cal G}_{bc}\, ({\cal G}_{dp}\varphi^p_2) \nonumber
\eea
\end{appendix}
\nocite{*}
\bibliography{fNL}
\bibliographystyle{utphys}


\end{document}